\RequirePackage{fix-cm}
\documentclass[smallcondensed]{svjour3}     % onecolumn (ditto)
\smartqed  % flush right qed marks, e.g. at end of proof
\usepackage{graphicx}
\usepackage{multirow}
\usepackage{natbib}
\usepackage{marginnote}
\usepackage[colorlinks=true,linkcolor=blue]{hyperref}
\usepackage{amsmath}
\usepackage{amssymb}    
\begin{document}

\title{Observable quality assessment of broadband very long baseline interferometry system%\thanks{Grants or other notes
%about the article that should go on the front page should be
%placed here. General acknowledgments should be placed at the end of the article.}
}
%\subtitle{Do you have a subtitle?\\ If so, write it here}

\titlerunning{Observable quality assessment of VGOS}        % if too long for running head

\author{Ming H. Xu \and James M. Anderson \and Robert Heinkelmann \and Susanne Lunz \and Harald Schuh \and Guangli Wang %etc.
}

%\authorrunning{Short form of author list} % if too long for running head

\institute{Ming H. Xu \at
              Aalto University Mets\"{a}hovi Radio Observatory, Mets\"{a}hovintie 114, 02540 Kylm\"{a}l\"{a}, Finland; Aalto University Department of Electronics and Nanoengineering, PL15500, FI-00076 Aalto, Finland; Technische Universit\"{a}t Berlin, Institut f\"{u}r Geod\"{a}sie und Geoinformationstechnik, Fakult\"{a}t VI, Sekr. KAI 2-2, Kaiserin-Augusta-Allee 104-106, D-10553 Berlin, Germany; Shanghai Astronomical Observatory, Chinese Academy of Sciences, 200030 Shanghai, China \\
%              Tel.: +123-45-678910\\
%              Fax: +123-45-678910\\
              \email{minghui.xu@aalto.fi}           %  \\
%             \emph{} of F. Author  %  if needed
           \and
           James M. Anderson \at 
           Technische Universit\"{a}t Berlin, Institut f\"{u}r Geod\"{a}sie und Geoinformationstechnik, Fakult\"{a}t VI, Sekr. KAI 2-2, Kaiserin-Augusta-Allee 104-106, D-10553 Berlin, Germany; Helmholtz Centre Potsdam, GFZ German Research Centre for Geosciences, Telegrafenberg, 14473 Potsdam, Germany
           \and
           Robert Heinkelmann, Susanne Lunz \at
           Helmholtz Centre Potsdam, GFZ German Research Centre for Geosciences, Telegrafenberg, 14473 Potsdam,
           Germany
           \and
           Harald Schuh \at
           Helmholtz Centre Potsdam, GFZ German Research Centre for Geosciences, Telegrafenberg, 14473 Potsdam,
           Germany; Technische Universit\"{a}t Berlin, Institut f\"{u}r Geod\"{a}sie und Geoinformationstechnik, Fakult\"{a}t VI, Sekr. KAI 2-2, Kaiserin-Augusta-Allee 104-106, D-10553 Berlin, Germany
           \and
           Guangli Wang \at
           Shanghai Astronomical Observatory, Chinese Academy of Sciences, 200030 Shanghai, China
}

%
%\institute{Ming H. Xu \at
%   first address \\
%  %            Shanghai Astronomical Observatory, Chinese Academy of Sciences, 200030 Shanghai, China\\
%                            Tel.: +123-45-678910\\
%              Fax: +123-45-678910\\
%                            \email{mhxu@shao.ac.cn}           %  \\
%%             \emph{Present address:} of F. Author  %  if needed
%           \and
%           James M. Anderson \at
%           Institute of Geodesy and Geoinformation Science, Technische Universit\"{a}t Berlin, Stra${\beta}$e des 17. Juni 135, 10623, Berlin, Germany
%%           \and
%           Robert Henkelmann, Sunsanne Lunx \at
%           Helmholtz Centre Potsdam, GFZ German Research Centre for Geosciences, Telegrafenberg 14473 Potsdam, Germany
%           \and
%           Harald Schuh \at
%           Institute of Geodesy and Geoinformation Science, Technische Universit\"{a}t Berlin, Stra${\beta}$e des 17. Juni 135, 10623, Berlin, Germany
%           \and
%           Guangli Wang \at
%           Shanghai Astronomical Observatory, Chinese Academy of Sciences, 200030 Shanghai, China
%
%}

\date{Received: 29 Nov. 2019 / Accepted: Feb. 2021}
% The correct dates will be entered by the editor

\maketitle

\begin{abstract}
The next-generation, broadband geodetic very long baseline interferometry
system, named VGOS, is developing its global network, and VGOS networks with
a small size of 3--7 stations have already made broadband observations from
2017 to 2019. We made quality assessments for two kinds of observables in
the 21 VGOS sessions currently available: group delay and differential total
electron content ($\delta$TEC). Our study reveals that the random
measurement noise of VGOS group delays is at the level of less than 2\,ps (1\,ps\,=\,10$^{-12}$\,s), while the contributions from systematic error
sources, mainly source structure related, are at the level of 20\,ps. Due to
the significant improvement in measurement noise, source structure effects
with relatively small magnitudes that are not overwhelming in the S/X VLBI
system, for instance 10\,ps, are clearly visible in VGOS observations. 
Another critical error source in VGOS observations is discrete delay jumps, 
for instance, a systematic offset of about 310\,ps or integer multiples of that. 
The predominant causative factor is found to be related to source structure. 
The measurement noise level of $\delta$TEC observables is about 0.07\,TECU, but the
systematic effects are five times larger than that. A strong correlation
between group delay and $\delta$TEC observables is discovered with a trend of
40\,ps/TECU for observations with large structure effects; there is a second trend
in the range 60\,ps/TECU to 70\,ps/TECU when the measurement noise is dominant.

% \PACS{PACS code1 \and PACS code2 \and more}
% \subclass{MSC code1 \and MSC code2 \and more}
\end{abstract}

\section{Introduction}
\label{intro}

Geodetic very long baseline interferometry (VLBI) is a space-geodetic
technique that has regularly made global astrometric/geodetic observations
since 1979, which are the basis for creating the International Celestial
Reference Frame \citep[ICRF2;][]{fey2015aj} and obtaining a full set of Earth
Orientation Parameters. Together with the other three space geodetic techniques,
VLBI plays an important role in establishing the International Terrestrial
Reference Frame \citep[ITRF2014;][]{altamimi2016jgr}. At the beginning of this
century, the International VLBI Service for Geodesy and Astrometry
\citep[IVS\footnote{https://ivscc.gsfc.nasa.gov/index.html};][]{schuh2012jg,nothnagel2017jg}
proposed to develop the next-generation geodetic VLBI system, initially called VLBI2010 \citep{niell2006} but subsequently renamed the VLBI Global Observing System (VGOS). This new VLBI system relies mainly on the
advantages of small ($\sim$12\,meters in diameter) and fast-slewing
antennas, ultra-wide observing frequency receivers (from 2\,GHz to\,14\,GHz), and the
expectation of continuous operation, 24\,hours a day and seven days a week
\citep{petrachenko2009}. In order to achieve its goal of 1\,mm position
accuracy and 0.1\,mm/yr velocity stability on global scales, the first strategy
proposed by the VGOS working group was to reduce the random noise component of
the group delays \citep{niell2007}. Building a global VGOS network with a
sufficient number of stations is in progress, and a small VGOS network has
started to make broadband observations. The technical implementation of the
VGOS system can be found in \citet{niell2018rs}, and the data correlation and
processing of VGOS observations from a single baseline can be referred to in
\citet{kondo2016rs} and \citet{niell2018rs}. Analyzing these actual VGOS
observations allows us to investigate the measurement noise level and the 
systematic behaviors of the VGOS observations.

In this paper we investigate the contribution of random measurement noise and
systematic error sources in VGOS delay and differential total electron content
($\delta$TEC) observables\footnote{An observable refers to a specific kind of quantity, such as amplitude, phase, delay or rate, that has been measured by maximizing the correlation between the recorded signals of a distant radio source at the two stations of a baseline; in addition, $\delta$TEC estimate is included as another kind of observable in VGOS.}. The relationship between these two types of
observables is also studied. We use a different method of assessing the
error level of the VGOS system than that in \cite{elosegui2018agu}
and \citet{niell2018rs}, who demonstrated the post-fit residuals from geodetic
VLBI solutions. Furthermore, instead of studying observations of one short
baseline, we present the results of VGOS observations from a global network.
In section 2 we present the VGOS observations currently available and
introduce the method of data analysis that we used. A quality assessment of
group delay observables is given in section\,3. In section\,4 we demonstrate
the measurement noise level and systematic errors in $\delta$TEC observables,
estimated simultaneously in VGOS observations. The strong correlation between
VGOS group delay and $\delta$TEC observables is studied in section\,5. In
section\,6 we summarize and discuss the results.

\section{Broadband VLBI observations and data analysis}
\label{section_data}
The IVS conducted a
continuous observing campaign with three VLBI networks (two legacy S/X 
networks and one VGOS broadband network) in 2017, 
called CONT17 \citep{2020JGeod..94..100B}. The VGOS broadband
network in CONT17 had a smaller number of stations than the two legacy
networks, and it observed only for one third of the whole CONT17 period.
However, it provides the first public data set of the VGOS broadband system,
which was originally proposed about 20 years ago. As of 15th Nov. 2019, 16
other VGOS sessions carried out in 2019 were released\footnote{Data
are available through the NASA CDDIS server:
https://cddis.nasa.gov/archive/vlbi/ivsdata/vgosdb/}, as listed in Table
\ref{tab:1}. On average, 24-hour VGOS sessions obtain about 2.2 times as many as 
scans\footnote{A scan consists of simultaneous observations of a radio source by two or more stations over an interval  on the order of 5\,seconds to 2\,minutes.} 
than the legacy 24-hour VLBI sessions. These broadband observations were
made simultaneously at four 512-MHz-wide bands centered at 3.2, 5.5, 6.6 and 10.4
GHz. (The detailed technical description of the observing frequency setup
is available in \citet{niell2018rs}.) The median and mean of the formal errors for group delay and
$\delta$TEC observables in each session are also shown in the table. The
median formal errors of group delay observables for these 21 sessions are in
the range of 1.2\,ps (1\,ps\,=\,10$^{-12}$\,s) to 3.0\,ps, and those for $\delta$TEC observables
are in the range of 0.029\,TECU (1\,TECU\,=\,10$^{16}$\,electrons per square meter) to 0.060\,TECU.

We processed the 21 VGOS sessions to determine error contributions in group
delay observables, including measurement noise and source structure effects,
by doing closure analysis \citep{xu2016aj, xu2017jg, anderson2018jgr}. 
We adopted the same procedure of closure analysis for the VGOS sessions as was developed for the CONT14 sessions described in \citet{anderson2018jgr}. (The
technical description of our closure analysis can be found in the supplemental
information to \citet{anderson2018jgr}.) In short, the method of closure
analysis statistically determines the baseline equivalent delay error of 
each individual observation\footnote{In the remainder of this paper, ``observation'' is used with the restricted meaning of a pair of two stations --- a baseline --- observing a radio source over a short duration, typically on the order of 5\,seconds to 2\,minutes.} from all the available closure
delays involving that observation, called closure-based error estimate; 
the weighted root-mean-square (WRMS) delay error of a group of data can then be derived by
combining the closure-based error estimates of the delay observables in the group. The method has two major advantages: (1) the station-based errors\footnote{We refer to effects such as atmosphere, ionosphere, clock, and geometry as station-based---when there is a change at one epoch for a station for any of these effects the corresponding changes with the same magnitude will happen to all the observations on the baselines of that station within the scan of that epoch---and the errors in modeling these effects as station-based errors.}
are canceled out exactly in closure delays; and (2) complementary to the post-fit residuals
from geodetic solutions, it provides an independent way of assessing the
observable quality. Except for baseline clocks, which in some cases are included in the parameterization as a constant offset in delays for a specific baseline and can thus only reduce a constant offset in closure delays, no geodetic parameters in a routine VLBI
solution can absorb nonzero closure delays. They therefore contribute entirely to the residuals of the VLBI solution and can bias the estimates of geodetic parameters. In the recent research of \citet{2019evga.conf..224B}, structure model parameters were included in the VLBI solution of the CONT17 VGOS sessions to reduce the large residual delays of the sources 0552$+$398 and 2229$+$695, which can thus reduce the magnitudes of the delay misclosures. However, the method has not been demonstrated to be applicable to general cases of radio sources with structure at different scales or insufficient numbers of observations. In the paper, closure delays, measuring intrinsic structure of sources as closure phases and closure amplitudes, are treated as errors in VGOS broadband delays only because the effects 
of source structure bias the geodetic parameters.

Closure analysis was also applied to the estimated 
ionosphere-like phase dispersion parameter, called $\delta$TEC, from these VGOS sessions. 
$\delta$TEC is the difference of the total electron content (TEC) along the line of sight from a source to each station of a baseline during a scan. Closure $\delta$TEC over a triangle of three antennas therefore gives insight into the errors in $\delta$TEC measurements.

The conditions for the exclusion of an observation, called flagging, are summarized here: (1) observations with signal-to-noise ratio (SNR) less than 7; (2) station \texttt{RAEGYEB} from the second day to the last
day of CONT17 VGOS observations, that is the sessions B17338, B17339, B17340 and B17341; and (3) all the observations on the baseline \texttt{ONSA13NE}--\texttt{ONSA13SW}.

\begin{table}[tbhp!]
% table caption is above the table
\caption{Observing sessions of the VGOS broadband network}
\label{tab:1}       % Give a unique label
% For LaTeX tables use
\begin{tabular}{lcrrclrrrr}
\hline\noalign{\smallskip}
Date & Session & Number & Number of& Number &  Station list& \multicolumn{2}{c}{Delay formal err.} & \multicolumn{2}{c}{$\delta$TEC formal err. }\\
(yyyy/mm/dd) & name &of scans & observations &of sources &  & Median & Mean & Median & Mean \\
(1) & (2) & (3) &(4)&(5)&(6)&\multicolumn{2}{c}{[ps]} & \multicolumn{2}{c}{[TECU]}\\
\noalign{\smallskip}\hline\noalign{\smallskip}
2017/12/03 & B17337 & 1180 & 5999 & 67 & GsIsK2YjWfWs     & 1.71 & 2.12 & 0.039 & 0.046 \\
2017/12/04 & B17338 & 1170 & 5037 & 66 & GsIsK2YjWfWs     & 3.01 & 5.00 & 0.060 & 0.098 \\
2017/12/05 & B17339 & 1180 & 5833 & 65 & GsIsK2YjWfWs     & 2.86 & 4.86 & 0.057 & 0.095 \\
2017/12/06 & B17340 & 1130 & 5166 & 66 & GsIsK2YjWfWs     & 2.41 & 4.50 & 0.051 & 0.089 \\
2017/12/07 & B17341 & 1246 & 6043 & 66 & GsIsK2YjWfWs     & 2.43 & 4.47 & 0.050 & 0.088 \\
2019/01/07 & VT9007 & 1132 & 8310 & 64 & GsK2OeOwYjWfWs   & 1.70 & 2.37 & 0.039 & 0.049 \\
2019/01/22 & VT9022 & 1024 & 6070 & 64 & K2OeOwYjWfWs     & 1.45 & 2.00 & 0.035 & 0.043 \\
2019/02/04 & VT9035 & 1043 & 4622 & 64 & GsK2OeYjWfWs     & 1.39 & 1.72 & 0.036 & 0.042 \\
2019/02/19 & VT9050 & 1115 & 7668 & 62 & GsK2OeYjWfWs     & 1.37 & 1.79 & 0.035 & 0.042 \\
2019/03/04 & VT9063 & 1129 & 7645 & 63 & GsK2OeYjWfWs     & 1.34 & 1.76 & 0.033 & 0.041 \\
2019/03/18 & VT9077 & 1080 & 5586 & 61 & GsK2OeYjWfWs     & 1.29 & 1.77 & 0.032 & 0.041 \\
2019/04/01 & VT9091 & 1121 & 7651 & 62 & GsK2OeYjWfWs     & 1.43 & 1.83 & 0.034 & 0.042 \\
2019/04/15 & VT9105 & 1105 & 5102 & 61 & GsK2OeWfWs       & 1.44 & 1.85 & 0.035 & 0.043 \\
2019/04/29 & VT9119 & 1126 & 5142 & 63 & GsK2OeWfWs       & 1.72 & 2.19 & 0.040 & 0.048 \\
2019/05/13 & VT9133 & 1123 & 4120 & 63 & GsK2OeWfWs       & 1.71 & 2.25 & 0.038 & 0.048 \\
2019/05/28 & VT9148 & 676 & 1444 & 60 & GsOeWs            & 1.21 & 1.67 & 0.031 & 0.039 \\
2019/06/11 & VT9162 & 1125 & 4891 & 64 & GsK2OeWfWs       & 1.43 & 1.84 & 0.034 & 0.041 \\
2019/06/24 & VT9175 & 1110 & 5097 & 66 & GsK2OeWfWs       & 1.71 & 2.20 & 0.039 & 0.047 \\
2019/07/08 & VT9189 & 776 & 1860 & 60 & GsOeWs            & 1.17 & 1.64 & 0.029 & 0.037 \\
2019/07/22 & VT9203 & 1093 & 6235 & 67 & GsK2OeOwWfWs     & 1.94 & 2.44 & 0.042 & 0.050 \\
2019/08/05 & VT9217 & 1174 & 11541 & 74 & GsK2OeOwYjWfWs  & 1.63 & 2.29 & 0.039 & 0.050 \\
%2019/08/19 &VT9231& 1213 & 12002 & 80 &GsK2OeOwYjWfWs& 1.53 & 1.96 & 0.036 & 0.043 \\
%2019/09/05 &VT9248& 1247 & 12584 & 63 &GsK2OeOwYjWfWs& 1.39 & 1.85 & 0.033 & 0.040 \\
%2019/09/16 &VT9259& 1238 & 12468 & 66 &GsK2OeOwYjWfWs& 1.48 & 1.97 & 0.035 & 0.042 \\
%2019/09/30 &VT9273& 1215 & 7143 & 69 &GsK2OeOwWfWs& 1.59 & 1.97 & 0.036 & 0.043 \\
%2019/10/17 &VT9290& 1030 & 3768 & 65 &GsK2OeOwWs& 1.43 & 1.83 & 0.035 & 0.042 \\
%2019/10/28 &VT9301& 1109 & 5154 & 56 &GsK2OeOwWf& 1.99 & 2.62 & 0.045 & 0.057 \\
%2019/11/14 &VT9318& 1097 & 3964 & 57 &GsK2OeOwYjWfWs& 1.84 & 2.43 & 0.043 & 0.055 \\
%2019/11/25 &VT9329& 1653 & 10733 & 92 &GsK2OeOwWfWs& 1.71 & 2.07 & 0.043 & 0.051 \\
%2019/12/09 &VT9343& 1175 & 8621 & 66 &GsK2OeOwWfWs& 1.66 & 2.06 & 0.041 & 0.049 \\
%2019/12/26 &VT9360& 1290 & 6365 & 81 &GsK2OeOwWfWs& 1.79 & 2.47 & 0.044 & 0.061 \\
%2020/01/09 &VO0009& 1173 & 8168 & 71 &GsIsOeOwWs& 1.61 & 2.11 & 0.040 & 0.048 \\
%2020/01/21 &VO0021& 1328 & 8256 & 69 &GsIsK2MgOeOwWs& 1.64 & 2.25 & 0.042 & 0.054 \\
%2020/02/03 &VO0034& 1637 & 13985 & 80 &GsIsK2MgOeOwWfWs& 2.00 & 2.90 & 0.047 & 0.066 \\
%2020/02/20 &VO0051& 1574 & 16987 & 79 &GsIsK2MgOeOwWfWsYj& 1.61 & 2.12 & 0.041 & 0.050 \\
%2020/03/02 &VO0062& 1585 & 13304 & 95 &GsK2MgOeOwWfWsYj& 1.86 & 2.43 & 0.047 & 0.057 \\
\noalign{\smallskip}\hline
\end{tabular}
\note{Two-letter station codes in column 6 have the following meanings:
Gs=\texttt{GGAO12M}, Is=\texttt{ISHIOKA}, K2=\texttt{KOKEE12M},
Yj=\texttt{REAGYEB}, Wf=\texttt{WESTFORD}, Ws=\texttt{WETTZ13S},
Oe=\texttt{ONSA12NE}, and Ow=\texttt{ONSA12SW}. Refer to 
ftp://cddis.gsfc.nasa.gov/pub/vlbi/ivscontrol/ns-codes.txt for more
information about these stations. The values in the last four columns are the median and mean of the formal errors for group delay and for the $\delta$TEC observables for observations with SNR $>$ 7. }
\end{table}

For completeness, we briefly recall the basic equations of the closure analysis and describe the terminology used. Closure delay is the sum of delay observables over a closed triangle of three stations. For a triangle of three stations, $a$, $b$, and $c$, closure delay is defined by
    \begin{equation}
     \label{eq_closure}
\tau_{\mbox{clr}}\equiv\tau_{ab}+\tau_{bc}+\tau_{ca},
    \end{equation}
where, for instance, $\tau_{ab}$ is the delay observable from station $a$ to
station $b$. The reference-time convention in geodetic VLBI defines that the timestamp
of the delay observable as the time of arrival of the wavefront at the first antenna of a baseline. For instance, delay $\tau_{ab}(t_0)$ refers to the delay for a wavefront that arrives at station $a$ at epoch of $t_0$. Therefore, the geodetic delay observables for multiple baselines in a scan, although they have the same timestamp, do not necessarily refer to the same wavefront. When these delay observables are used to derive closure delays, a correction is needed to make the geometry of a triangle completely close; detailed discussions and dedicated equations can be found in Section~2 of~\citet{xu2016aj} and in Section~4.1 of~\citet{anderson2018jgr}. An alternative way of forming closure delays is to use the delay observables with geocentric timestamps (the astronomical convention), rather than the delay observables used in geodetic solutions; the former need no correction. 

The uncertainty of a closure delay is calculated from the formal errors of the three observables forming it by assuming that they are independent.

For the delay observable $\tau_{ab}$ at a single epoch, its closure-based error estimate, ${\Delta}{\tau}_{ab}$, is statistically determined from all the closure delays that are formed by $\tau_{ab}$ together with the other un-flagged observations in the scan at that epoch, written as
%
%    \begin{equation}
%     \label{eq_closure_error_median}
%\Delta\tau_{ab}=\frac{\tau_{\mbox{clr}-ab}^{\left \lfloor{(N+1)/2}\right \rfloor} + \tau_{\mbox{clr}-ab}^{\left \lceil{(N+1)/2}\right \rceil}}{2\sqrt{3}},
%    \end{equation}

    \begin{equation}
     \label{eq_closure_error_mean}
\Delta\tau_{ab}=\frac{\sum^{N}_{i=1}  |\tau_{\mbox{\scriptsize clr}-ab}^{i}|}{\sqrt{3}N},
    \end{equation}
where $N$ is the number of such closure delays and $\tau_{\mbox{\scriptsize clr}-ab}^{i}$ is the $i$-th one. The number $\sqrt{3}$ in the denominator scales the mean closure delay to derive a baseline equivalent error by assuming that the errors in different observations are independent. This process as defined by equation \ref{eq_closure_error_mean} for the observable $\tau_{ab}$ was repeated for all observations one by one to derive their closure-based error estimates, $\Delta\tau$, whenever possible.

The WRMS delay error (not uncertainty), $\delta\tau$, is obtained by combining the closure-based error estimates as follows:

    \begin{equation}
     \label{eq_delay_error}
\delta\tau=\sqrt{\frac{\sum^{l}_{j=1}w_{j}(\Delta\tau^{j})^{2}}{\sum^{l}_{j=1}w_{j}}},
    \end{equation}
where $l$ is the number of un-flagged observations with closure-based error estimates available in a data group of interest (e.g., all observations of a particular source or some selected sources or all observations in one session), $\Delta\tau^{j}$ is the closure-based error estimate of the $j$-th observation, and $w_{j}$ is its weight. The weighting is done by setting an equal weight for all the delay observables, named uniform weighting, or by using the reciprocal of the square of the uncertainty (formal error) of each individual delay, named natural weighting. (The uniform and natural weighting schemes used here have different meanings to those used in the astrophysical imaging studies.) The same procedure of this closure analysis was applied to study $\delta$TEC measurements; closure $\delta$TEC, closure-based error estimate of $\delta$TEC and WRMS $\delta$TEC error are likewise defined.

Note that the closure analysis derives the \emph{baseline equivalent} error for each observation 
from closure quantities. It is obvious that the closure-based error estimate of an observation  
is affected (can be enlarged or reduced) by source structure effects and measurement noise in the observations of the other baselines in the scan. It is not appropriate to use closure-based error estimate to quantify the errors at the level of a single observation; however, the aim of closure analysis is to use closure-based error estimates 
only to determine the overall variance of source structure effects and measurement noise 
for a given group of data, as defined by equation \ref{eq_delay_error}. In this case, 
it will work without introducing significant biases when the random measurement noise, 
independent between different observations, is the dominant error
source. On the other hand, if the systematic error sources dominate, the mean of the absolute values of all the closures formed with a common observation maximizes the possibility of determining these systematic errors in that observation; 
it was then scaled by a factor of $\sqrt{3}$ to
reduce the contributions of systematic errors in the other observations forming those closures. Nevertheless, in the presence of systematic errors, 
the assumption for equation \ref{eq_closure_error_mean}
is not satisfied. It can lead to biases in interpreting the derived WRMS delay errors as the magnitudes of source structure effects that one would expect to have in the post-fit residuals from geodetic solutions. 
In order to investigate the potential biases, the median was used in place of the mean in Equation 
\ref{eq_closure_error_mean} as an alternative statistic to derive closure-based error estimates and the corresponding WRMS errors.

In closure analysis, we also directly compare the closure delays 
for a given source between various triangles 
and for a specific triangle between different sources, which can yield
insight into the properties of individual sources, baselines, and
stations.

%; it is also possible  that a closure delay is almost entirely due 
%to the systematic error of an observation, but its closure-based 
%error estimate is determined by reducing the closure delay by a 
%factor of $\sqrt{3}$
%
%For quantifying systematic error sources, it should work at least in the relative 
%sense because the magnitudes of closures entirely indicate the amount of errors in observations.
%Therefore, comparison can be made between closure analysis of different data groups, for example,
%from observations of different radio sources, different sessions or from VGOS observations and S/X observations.}}

\section{Group delay observables}
\subsection{Measurement noise}

A closure-based delay error estimate could be derived for 88\% of the 
observations in the 21 VGOS sessions, while 5\,$\%$ did not form any closure with un-flagged observations, and
7\,$\%$ were flagged as described in the previous section. Based on the uniform
weighting and the natural weighting schemes, the WRMS delay
error was calculated from these closure-based error estimates for each individual session
and for all 21 sessions combined. The results are shown in Table \ref{tab:2}. 
% For tables use
\begin{table}[tbhp!]
% table caption is above the table
\caption{WRMS delay errors determined by closure analysis (in units of picoseconds)}
\label{tab:2}       % Give a unique label
% For LaTeX tables use
\begin{tabular}{lrrrr}
\hline\noalign{\smallskip}
Session/Group & $N_{\texttt{obs}\ }$ & $N_{\texttt{CloErr}\ }$ & Uniform Weighting & Natural Weighting \\
(1) & (2) & (3) &(4) & (5)\\
\noalign{\smallskip}\hline\noalign{\smallskip}
B17337&  5999 &  5620 &       22.5 &       17.9\\
B17338&  5037 &  3279 &       26.8 &       19.1\\
B17339&  5833 &  3556 &       20.6 &       17.2\\
B17340&  5166 &  3042 &       24.4 &       20.3\\
B17341&  6043 &  3742 &       24.1 &       21.1\\
VT9007&  8310 &  7508 &       36.1 &       33.8\\
VT9022&  6070 &  5222 &       23.3 &       18.9\\
VT9035&  4622 &  4283 &       18.5 &       14.1\\
VT9050&  7668 &  7511 &       18.8 &       17.0\\
VT9063&  7645 &  7503 &       20.9 &       19.3\\
VT9077&  5586 &  5294 &       20.7 &       17.4\\
VT9091&  7651 &  7325 &       21.0 &       21.1\\
VT9105&  5102 &  4835 &       19.2 &       18.6\\
VT9119&  5142 &  4856 &       21.0 &       25.4\\
VT9133&  4120 &  3786 &       21.7 &       17.7\\
VT9148&  1444 &  1146 &       22.9 &       25.4\\
VT9162&  4891 &  4583 &       21.5 &       19.2\\
VT9175&  5097 &  4830 &       21.3 &       16.9\\
VT9189&  1860 &  1611 &       23.6 &       28.5\\
VT9203&  6235 &  5827 &       20.9 &       26.0\\
VT9217& 11541 & 11348 &       22.7 &       20.6\\
\textbf{ALL}& \textbf{121062} & \textbf{106707} &        \textbf{22.9} &     \textbf{21.0} \\
\textbf{ALL-19}& \textbf{106682} & \textbf{93977} &      \textbf{21.5} &       \textbf{20.0}\\
\textbf{CARMS-0.25} & \textbf{20998} & \textbf{17702} &  \textbf{6.2} &     \textbf{2.4} \\
\noalign{\smallskip}\hline
\end{tabular}{}
\note{$N_{\mbox{\scriptsize obs}}$ is the number of observations in each session or subgroup of data, and $N_{\mbox{\scriptsize CloErr}}$ is the number of observations that were not flagged out and formed at least one closure delay with un-flagged observations allowing the derivation of closure-based error estimates. }
\end{table}

Apart from session VT9007, the WRMS delay errors for the other 20 sessions are in
the range of 18.5\,ps to 26.8\,ps based on the uniform weighting and in the range
of 14.1\,ps to 28.5\,ps based on the natural weighting. The WRMS delay errors
for the 21 sessions combined, labelled as ``ALL" in the table, are about 23\,ps
and 21\,ps based on the two weighting schemes. This is a significant
improvement compared to the corresponding values of 35.3\,ps (uniform) and 25.2\,ps (natural)
for the CONT14 sessions\citep{anderson2018jgr}, 
which represent the best observing campaign of the legacy
S/X VLBI system.

\begin{figure*}[tbhp!]
% Use the relevant command to insert your figure file.
% For example, with the graphicx package use
  \includegraphics[width=0.75\textwidth]{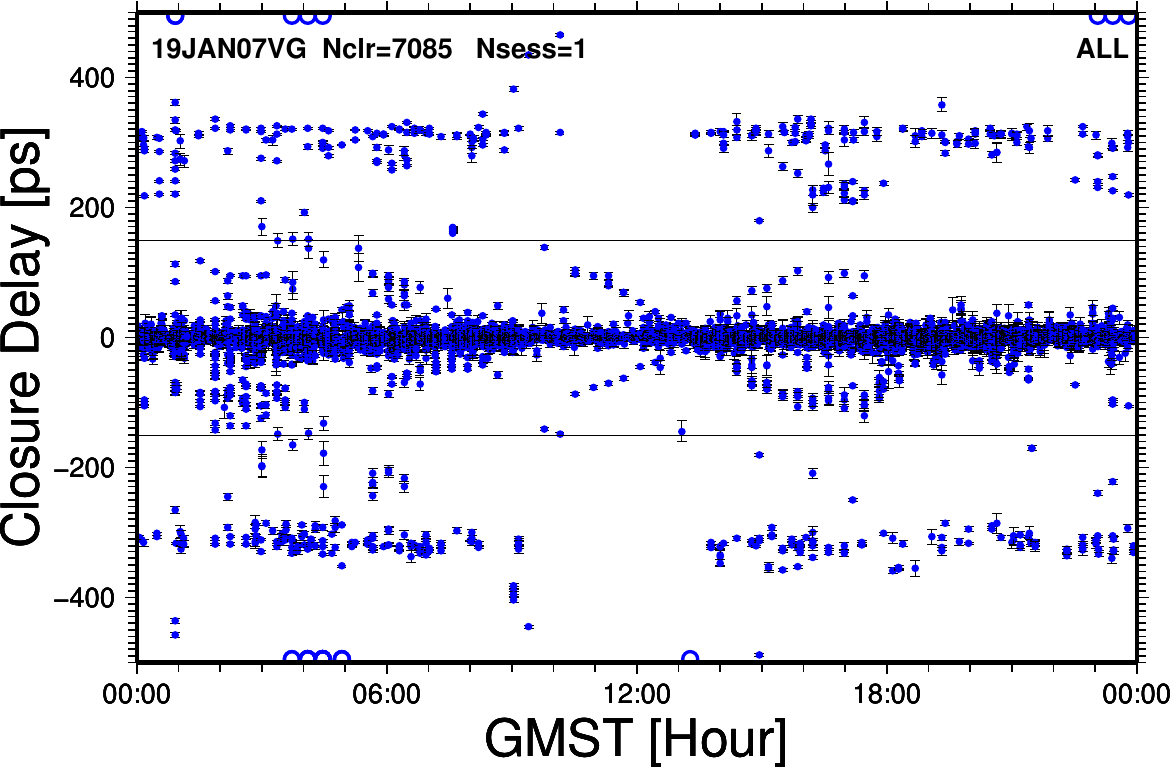}
% figure caption is below the figure
\caption{All closure delays of session VT9007 excluding triangles with baseline
\texttt{ONSA13NE}--\texttt{ONSA13SW}. Closure delay uncertainties
are shown as black bars. There are 7085 closure delays in total. A large number of
closure delays with an absolute offset of about 310\,ps is visible. All the closure delays
exceeding the limits of the $Y$ axis are shown on the top or bottom of the plot as open
circles. This convention applies to all of the closure plots in the paper; plots with no open-circle points on the bottom and top have no excessively large closure delays. Two solid horizontal lines with an absolute value of 150\,ps are provided as guides.}
\label{fig:19JAN07VG}       % Give a unique label
\end{figure*}

For session VT9007 the WRMS delay errors are remarkably high --- 36\,ps and 34\,ps from
the two weighting schemes. This is due to an exceptionally large number of 
misclosures of about 310\,ps or $-$310\,ps in the closure delays, as shown in Fig. \ref{fig:19JAN07VG}. 
The vast majority of these misclosures involve station $\texttt{ONSA13SW}$, due to its phasecal problem at the 
6.6-GHz frequency band (Brian Corey, personal communication, September 7, 2020). After 390 closure delays of station $\texttt{ONSA13SW}$ with absolute values of about 310\,ps were flagged, the WRMS delay error for session VT9007 was redetermined to be 23.0\,ps (uniform) and 18.9\,ps (natural). 
The WRMS delay errors for the ``ALL" group were recalculated
from the 19 sessions excluding VT9007 and VT9022---the latter session
undergoes the same issue but with offsets of around 1100\,ps and $-$1100\,ps, also 
related to station $\texttt{ONSA13SW}$, but not as many. 
The WRMS delay errors for the 19 sessions are 21.5\,ps (uniform) and 20.0\,ps (natural), 
labelled as ``ALL-19" group in Table \ref{tab:2}. In
summary, we argue that the magnitude of the random measurement noise and the systematic errors in 
the VGOS observations is in the range of 20.0\,ps to 22.9\,ps.

Except for sessions like VT9148 and VT9189 with an observing network of three stations, the natural
weighting scheme generally gives significantly smaller values of the WRMS delay error than the
uniform weighting scheme. This is to be expected when the non-Gaussian delay values due to source
structure are added to the closure delays with an otherwise noise-like distribution. 
On the other hand, because source structure effects not only cause structure delays in delay observables but also reduce observed amplitudes and thus the observations' SNR, natural weighting will underestimate the magnitude of their actual impacts. Thus, while the natural
weighting statistics are appropriate for evaluating the properties of the delay/$\delta$TEC observables, the
uniform weighting statistics can be useful for identifying sources with systematic errors, such as those
due to source structure. Furthermore, the SNRs of VGOS observations are typically very high, for instance, the median SNR for the CONT17 VGOS observations is $\sim$90; uniform weighting should be used to investigate the systematic error levels, especially if these systematic errors are significantly larger than the random measurement noise and are correlated with the SNRs, for example, source structure effects.

% For two-column wide figures use

\begin{figure*}[!htbp]
% Use the relevant command to insert your figure file.
% For example, with the graphicx package use
  \includegraphics[width=0.75\textwidth]{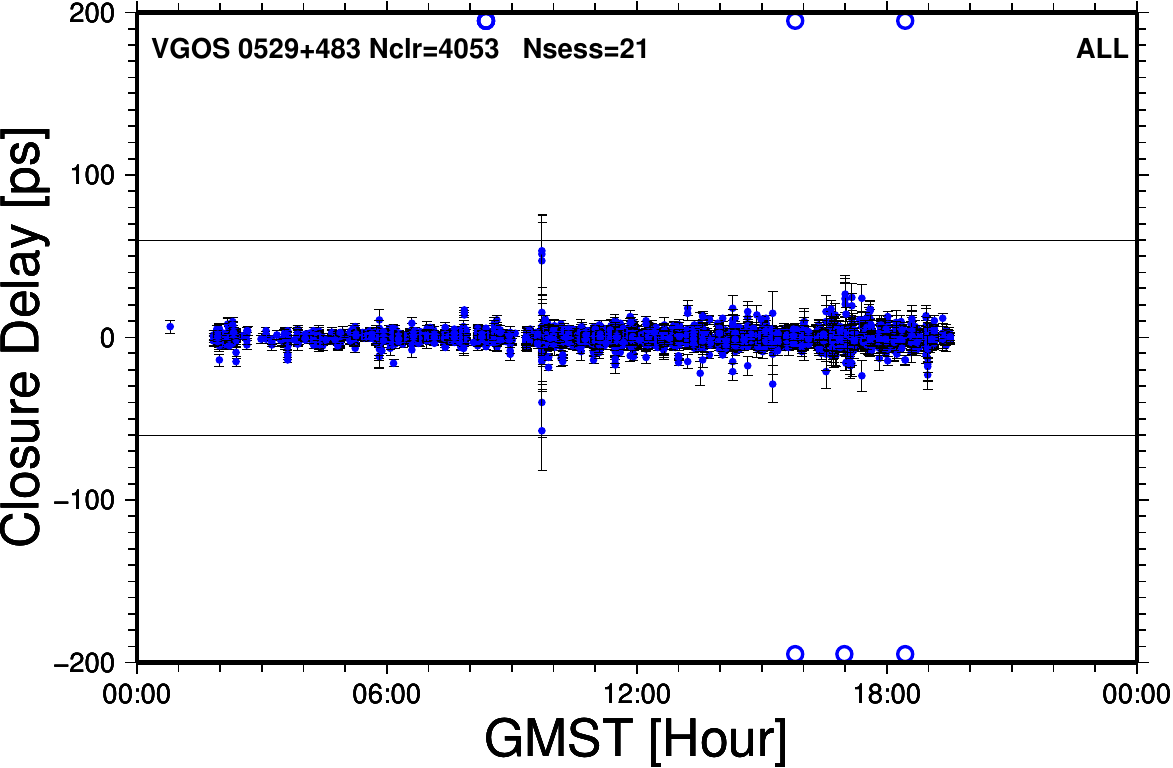}
% figure caption is below the figure
\caption{All closure delays of source 0529+483 in the 21 VGOS sessions with black bars giving
the 1-$\sigma$ measurement uncertainties based on the formal errors of delay observables.
There are four closure delays of about 310\,ps from one scan of session VT9007 showing as \texttt{one}
open circle on the top right of the figure and five closure delays of about 1100\,ps or $-$1100\,ps
from three scans of session VT9022 shown on the top and bottom of the figure.
The WRMS of all the available closure delays excluding these 9 is only 3.0\,ps from natural weighting. Source
0529+483 demonstrates the measurement noise level in VGOS delays, which should obviously be below 3\,ps. Two solid horizontal lines with an absolute value of~60\,ps are provided as guides.}
\label{fig:0529+483_all}       % Give a unique label
\end{figure*}

\begin{figure*}[!htbp]
  \includegraphics[width=0.75\textwidth]{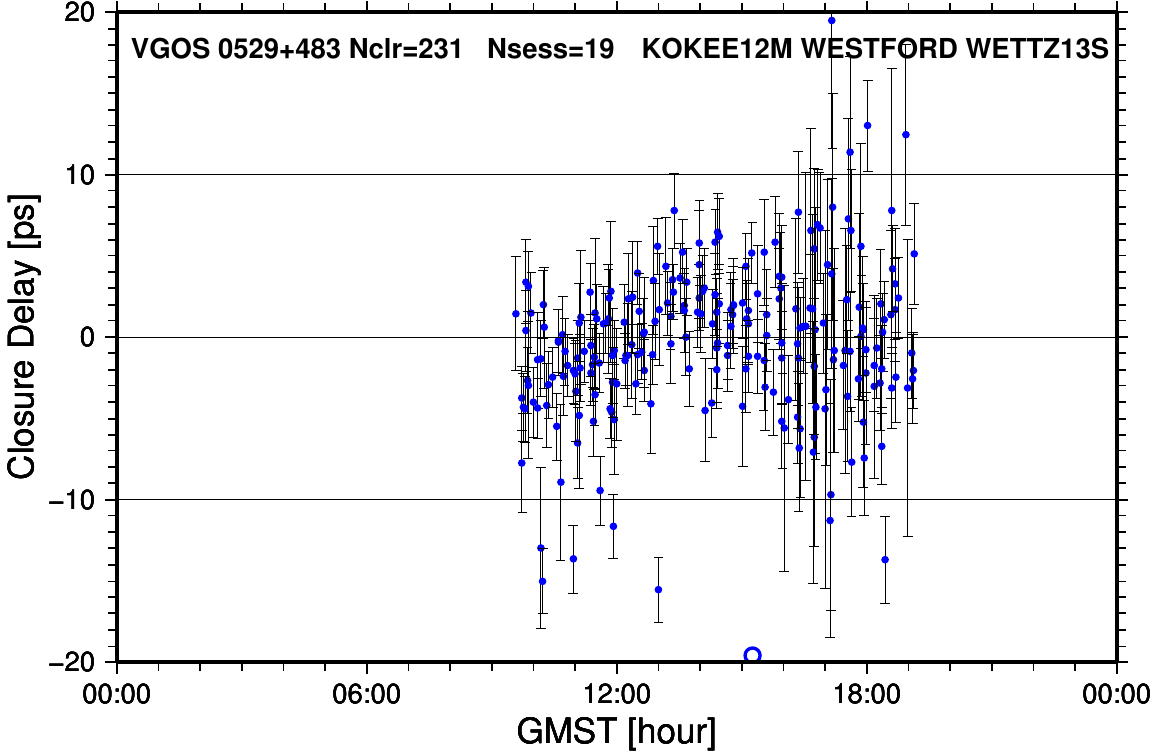}
% figure caption is below the figure
\caption{Zoom-in plot of closure delays of source 0529$+$483 for triangle \texttt{KOKEE12M}--\texttt{WESTFORD}--\texttt{WETTZ13S}. They are not randomly distributed around zero, suggesting that there are systematic effects with a magnitude of a few picoseconds for this source. Three solid horizontal lines are provided to guide the reader. }
\label{fig:0529+483}       % Give a unique label
\end{figure*}

In order to further investigate the random measurement noise level in VGOS
sessions, we adopted the closure amplitude RMS (CARMS) values based on the basic weighting scheme\footnote{It assumes that the noise floor in log closure amplitudes is 0.1 and thus adds 0.1 to their formal errors in the quadrature sense for weighting.} from Table 2 in \citet{xu2019apjs} to
identify the sources with minimum structure in these
VGOS sessions. For the definition of CARMS, please consult equations (2)--(4) and (6)--(8) in \citet{xu2019apjs}. The CARMS value of each individual source was calculated using all the available closure amplitudes for 
X-band only from historical VLBI
observations from 1980 to Aug. 2018 (no VGOS broadband observations are included). Apart from thermal noise, observations of an ideal point source will always give log closure amplitudes\footnote{Note that the natural logarithm was adopted in the definition of closure amplitude to calculate CARMS values, as shown in the equation (3) in \citet{xu2019apjs}.} equal to zero, while those of radio sources with extended structure will have log closure amplitudes deviating from zero, leading to larger CARMS values. Hence, in general, a smaller CARMS value of a source indicates that it causes less structure effects. Our recent study has demonstrated the correlation between the magnitudes of the radio-to-optical source position differences and CARMS values \citep{2021arXiv210112685X}. Using a maximum CARMS limit of 0.25 to
select sources with minimum structure, 28 low-structure sources were found in
the VGOS measurements, shown in Table \ref{tab:3}. 
The CARMS value of 0.25 was chosen as a compromise in order to have a sample of radio sources with both
minimum structure and a sufficient number of observations.
These 28 sources are associated
with 19.7 percent of the observations in the 19\,VGOS sessions (excluding sessions
VT9007 and VT9022). The WRMS delay error value for these observations, labelled
as ``CARMS-0.25" in Table~\ref{tab:2}, is 6.2\,ps for the uniform
weighting and only 2.4\,ps for the natural weighting. As we
explained already, the uniform weighting indicates the systematic error
contribution and the natural weighting tends to show the measurement
noise level. We therefore conclude that the VGOS measurement noise is no
larger than the 2\,ps level as demonstrated by the sources with minimum
structure, and the contributions of systematic errors for these sources are at
the level of 5\,ps to 6\,ps. Taking source 0529$+$483 as an example, all available closure delays in the 21 sessions are shown in Fig. 
\ref{fig:0529+483_all}. If the four closure delays in VT9007 with an
offset of 310\,ps and the five closure delays in VT9022 with offsets of 1100\,ps or
$-$1100\,ps are excluded, the WRMS closure delay for source 0529$+$483 is only
3.0\,ps. However, its closure delays, when inspecting one specific triangle at scales of a few tens of picoseconds, are still not randomly distributed, as shown in Fig. \ref{fig:0529+483}. 
Even though the magnitude of the systematic variations is only about 10\,ps, they are visible in the plot.
Similar or even larger systematic variations were detected for other CARMS-0.25 sources, such as 0716+714
and 0133+476.

% For tables use
\begin{table}[!htbp]
% table caption is above the table
\caption{Source group with CARMS less than 0.25, CARMS-0.25 for short}
\label{tab:3}       % Give a unique label
% For LaTeX tables use
\begin{tabular}{lcrc}
\hline\noalign{\smallskip}
IVS & CARMS & $N_{\texttt{obs}\ }$ & \ ICRF3 \\
Design. &  &  &category \\
(1) & (2) & (3) &(4)\\
\noalign{\smallskip}\hline\noalign{\smallskip}
0048$-$097 &   0.11 &  95    &  D  \\
0054$+$161 &   0.10 &  56    &  D  \\
0133$+$476 &   0.23 &  2906  &  D  \\
0237$-$027 &   0.15 &  193   &  D  \\
0446$+$112 &   0.24 &  589   &  O  \\
0529$+$483 &   0.21 &  3120  &  D  \\
0536$+$145 &   0.17 &  13    &  D  \\
0627$-$199 &   0.15 &  92    &  D  \\
0656$+$082 &   0.24 &  36    &  O  \\
0716$+$714 &   0.20 &  4865  &  D  \\
0723$+$219 &   0.18 &  13    &  O  \\
0727$-$115 &   0.24 &  727   &  D  \\
0804$+$499 &   0.20 &  211   &  D  \\
1040$+$244 &   0.17 &  815   &  D  \\
1124$-$186 &   0.21 &  312   &  D  \\
1243$-$160 &   0.13 &  288   &  D  \\
1300$+$580 &   0.18 &  1386  &  D  \\
1417$+$385 &   0.17 &  53    &  O  \\
1519$-$273 &   0.18 &  119   &  D  \\
1636$+$473 &   0.24 &  141   &  D  \\
1749$+$096 &   0.22 &  1163  &  D  \\
1908$-$201 &   0.20 &  222   &  D  \\
2059$+$034 &   0.24 &  33    &  D  \\
2141$+$175 &   0.21 &  730   &  O  \\
2215$+$150 &   0.22 &  1656  &  D  \\
2227$-$088 &   0.24 &  692   &  D  \\
2255$-$282 &   0.18 &   89   &  O  \\
2309$+$454 &   0.21 &  383   &  O  \\
\noalign{\smallskip}\hline
\end{tabular}
\note{The CARMS values in column 2 are taken from \citet{xu2019apjs},
and the ICRF3 categories in column 4
are from http://hpiers.obspm.fr/icrs-pc/newwww/icrf/icrf3sx.txt. D means defining sources, i.e., this particular source was included for the definition of the reference frame axes, and O means other non-defining sources. 
$N_{\mbox{obs}}$ in column 3 is 
the total number of VGOS observations in these 21 sessions for each source.}
\end{table}

% For two-column wide figures use

%
%In the cases of
%VLBI networks with more than 10 stations, RMS delay error value of
%``Median" indicates the lower boundary of error contributions, while that of
%``Mean" provides better error estimates (See Section B of the
%supplemental information of \citet{anderson2018jgr}). For a
%network in these 21 sessions, the differences in values from the two techniques become extremely small, no
%more than 0.5\,ps in most cases. The reason that the two techniques give similar
%results is that many triangles in the current VGOS network have similar geometries and 
%the maximum number of independent closure delays in one scan is $(n-1)(n-2)/2$, 
%where $n$ is the number of stations. A small number of such triangles available in each scan using different techniques to estimate delay errors does not make a significant difference.

As discussed at the end of section \ref{section_data}, the median value
was also used to derive closure-based error estimates and then to calculate the corresponding
WRMS delay errors. The differences in WRMS delay error values between the two techniques 
are very small for both weighting schemes, no more than 0.5\,ps in most cases.

\subsection{Source structure effects}

As we did for the historical S/X VLBI observations \citep{xu2019apjs}, it is
beneficial to show a few closure plots for several sources with different
magnitudes of structure effects as examples to understand those
effects in the broadband VLBI system.

\begin{figure*}[!htbp]
  \includegraphics[width=0.75\textwidth]{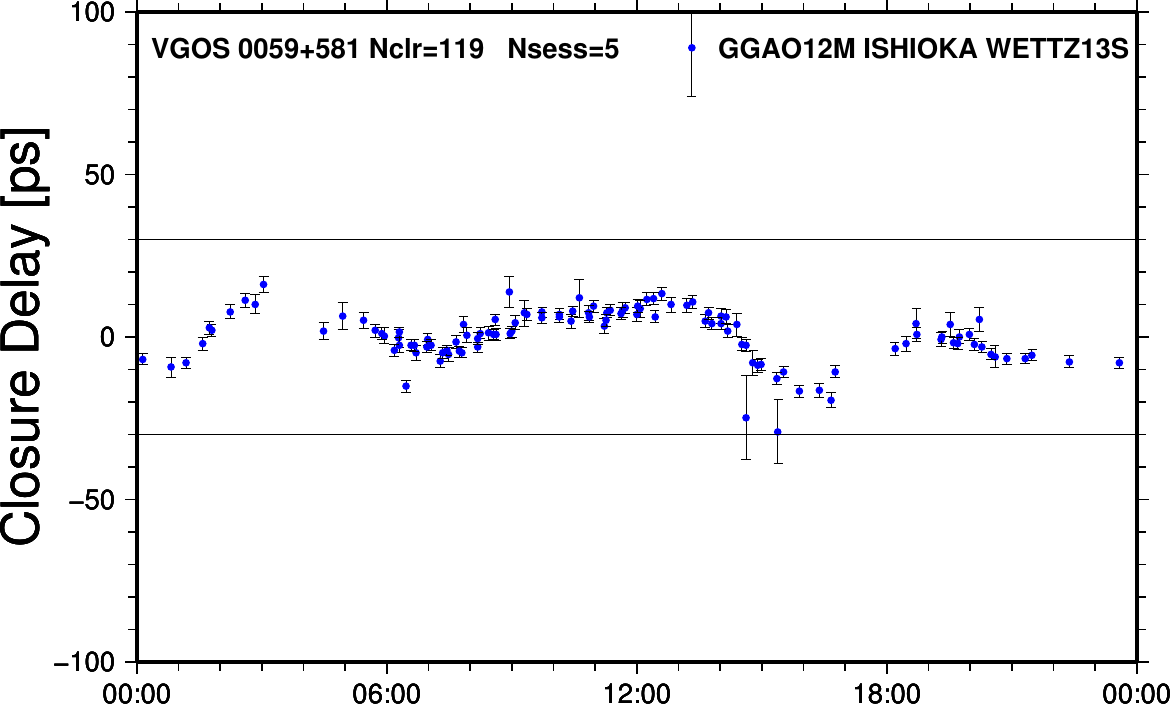}
  \includegraphics[width=0.75\textwidth]{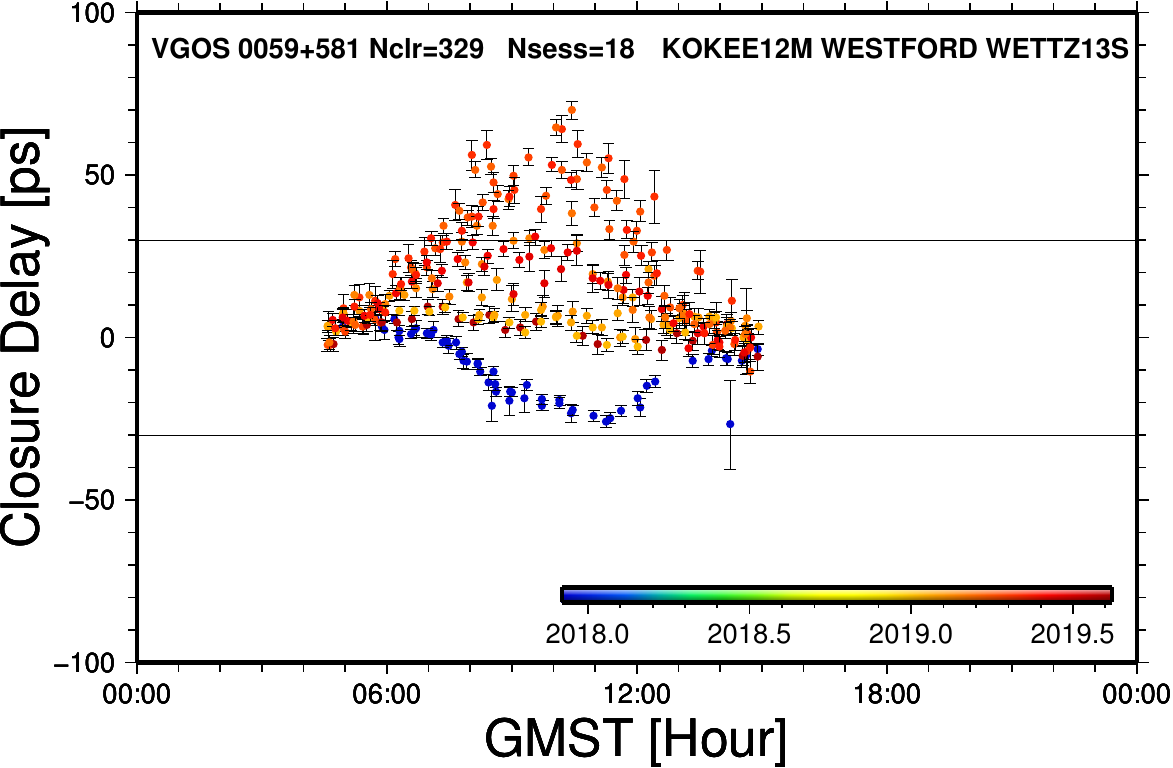}
% figure caption is below the figure
\caption{Closure delays for source 0059$+$581 as a function of GMST for two triangles,
\texttt{GGAO12M}--\texttt{ISHIOKA}--\texttt{WETTZ13S} (top) and
\texttt{KOKEE12M}--\texttt{WESTFORD}--\texttt{WETTZ13S} (bottom). The color coding indicates the
observation date, and the corresponding legend is shown
on the bottom-right corner of the bottom plot. The top plot shows a normal pattern
of source structure effects, while the bottom one clearly shows the source-structure time evolution
from CONT17 in Dec. 2017 to 2019 and even within 2019. Two solid horizontal lines with an absolute value of 30\,ps are provided as guides.}
\label{fig:0059+581}       % Give a unique label
\end{figure*}

\paragraph{0059$+$581} Closure delay plots for source 0059$+$581 are shown in Fig.
\ref{fig:0059+581} for two triangles,
\texttt{GGAO12M}--\texttt{ISHIOKA}--\texttt{WETTZ13S} and
\texttt{KOKEE12M}--\texttt{WESTFORD}--\texttt{WETTZ13S}. The first triangle
was observed only in CONT17 and has 119 closure delays in total. The pattern
of two peaks with opposite signs separated by a 12-hour GMST period is a
normal behavior of source structure effects. The second triangle, which was observed in
18\,VGOS sessions, produced 329 closure delays. Through it, the
source-structure time evolution is well demonstrated: the peak in the closure
delay pattern changed from $-$30\,ps in Dec. 2017 to around 0\,ps in early 2019,
increased to $+$60\,ps in March and decreased back to $+$30\,ps in the middle of
2019. Source 0059$+$581 is a very typical geodetic source and has been the
most frequently observed source both by the legacy VLBI system and the VGOS
system so far. For the triangle
\texttt{GGAO12M}--\texttt{ISHIOKA}--\texttt{WETTZ13S}, it is seen that the
structure effects have a magnitude of as large as 20\,ps but the WRMS
closure delay is only 6.9\,ps. Source structure effects are more easily
visible in VGOS observations than in the legacy VLBI observations
because the measurement noise in VGOS is well below 3\,ps. This is one reason
why source structure effects are so critical for VGOS.

\begin{figure*}[!htbp]
  \includegraphics[width=0.75\textwidth]{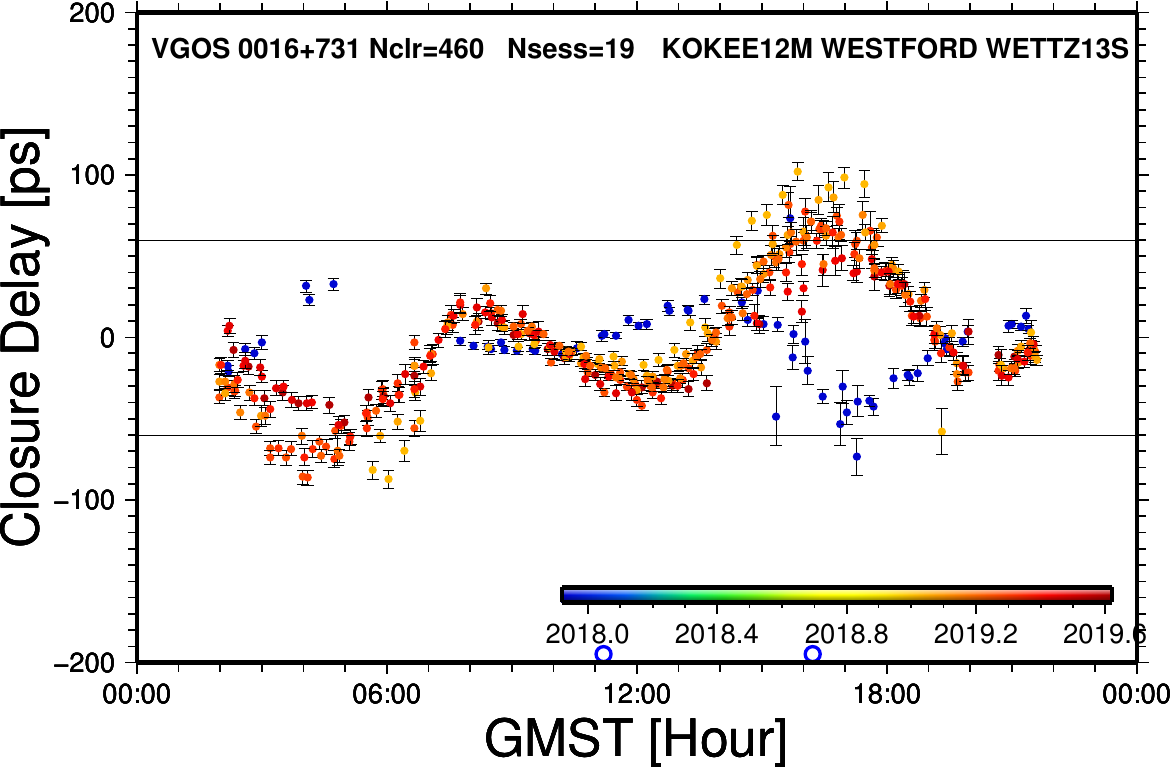}
%  \includegraphics[width=0.75\textwidth]{observations/all_delay_VGOS_0016+731_GGAO12M.pdf}
% figure caption is below the figure
\caption{Plot of closure delays for source 0016$+$731 as a function of GMST for triangle
\texttt{KOKEE12M}--\texttt{WESTFORD}--\texttt{WETTZ13S}, which was shown also for source 0529+483 in Fig. \ref{fig:0529+483} and 
for source 0059$+$581 in the bottom of Fig. \ref{fig:0059+581}. Source 0016$+$731 is another one of
the important geodetic sources. However, its structure effects have significantly larger
amplitudes than those of source 0059$+$581. Two solid horizontal lines with an absolute value of 60\,ps are provided as guides.}
\label{fig:0016+731}       % Give a unique label
\end{figure*}

\paragraph{0016$+$731} Source 0016$+$731 is another of
the important geodetic sources. The closure delays for source 0016$+$731 are
shown in Fig. \ref{fig:0016+731} for triangle
\texttt{KOKEE12M}--\texttt{WESTFORD}--\texttt{WETTZ13S}, which is the same
triangle shown in Fig. \ref{fig:0529+483} for source 0529+483 and in the bottom plot of Fig. \ref{fig:0059+581} for source 0059$+$581. It has 460\,closure
delays in 19 VGOS sessions. The source structure changed significantly from
2017 to 2019. The magnitudes of structure effects are as large as 100\,ps in
2019.

\begin{figure}[tbhp!]
% Use the relevant command to insert your figure file.
% For example, with the graphicx package use
  \includegraphics[width=0.75\textwidth]{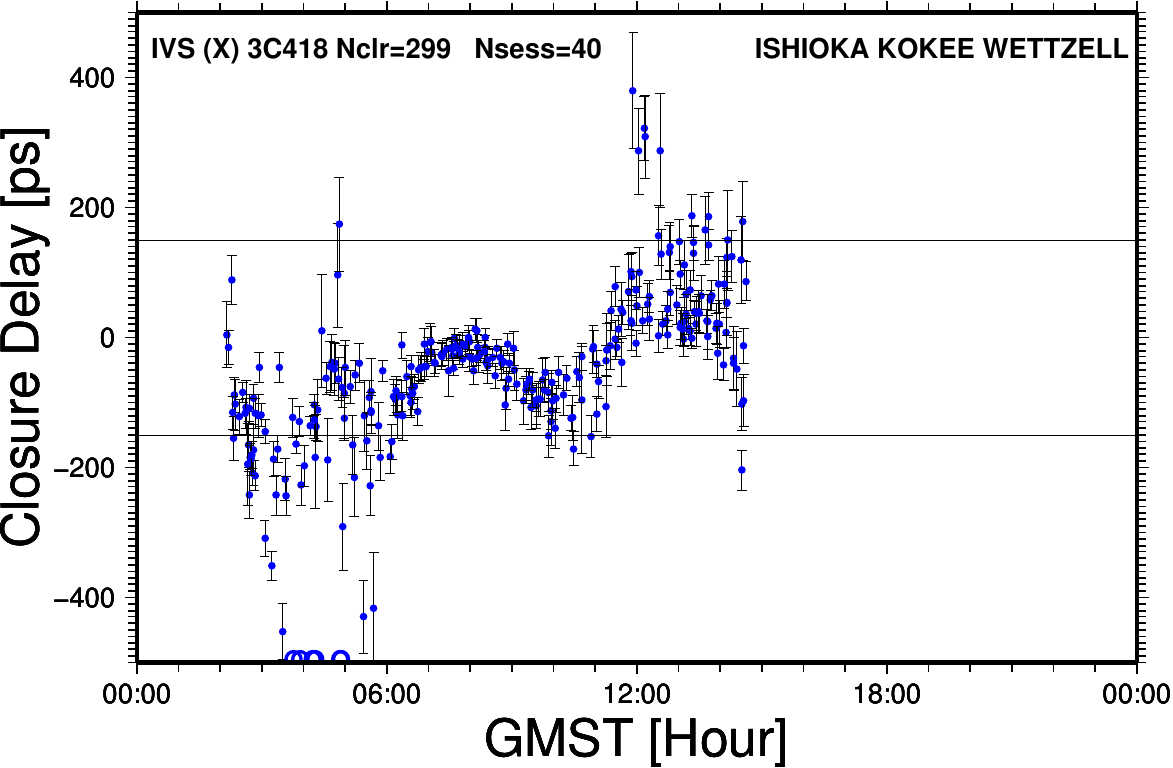}
  \includegraphics[width=0.75\textwidth]{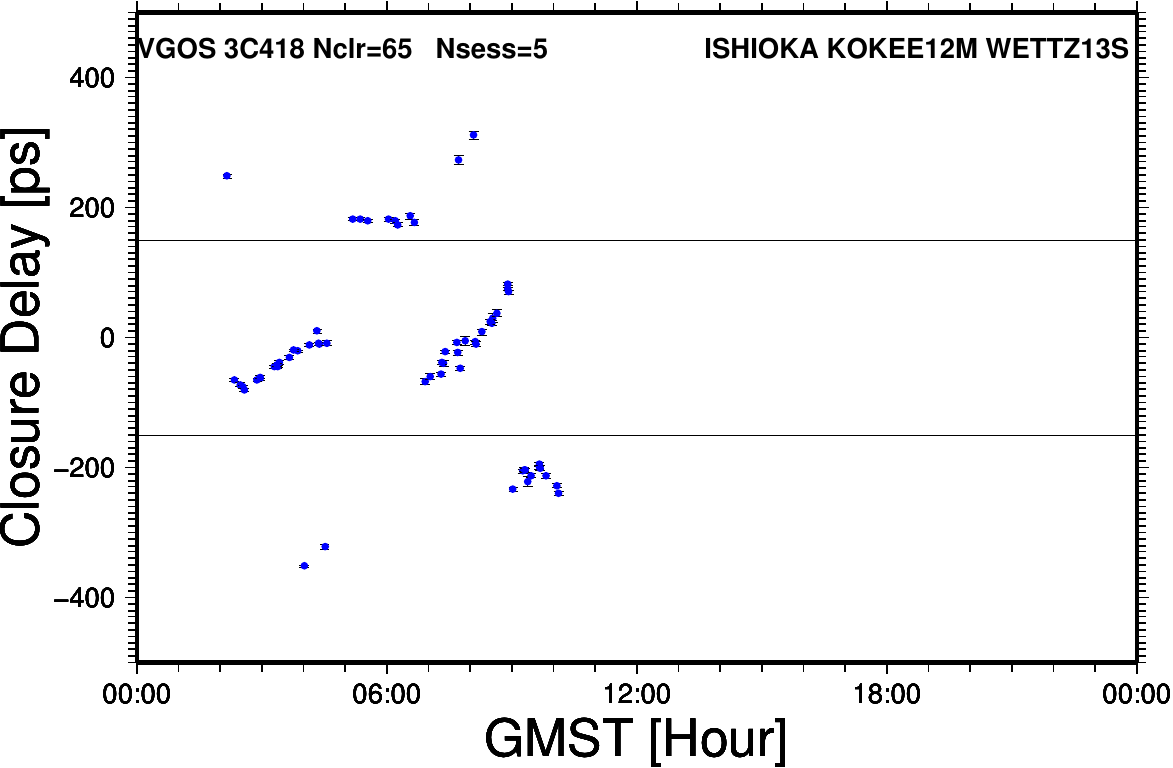}
% figure caption is below the figure
\caption{Plots of closure delays for source 3C418 as a function of GMST for two triangles,
\texttt{ISHIOKA}--\texttt{KOKEE}--\texttt{WETTZELL} (top, legacy X-band) and
\texttt{ISHIOKA}--\texttt{KOKEE12M}--\texttt{WETTZ13S} (bottom, VGOS). With replaceable S/X and
broadband receivers at station \texttt{ISHIOKA}, the first triangle observed in the S/X mode while
the second one observed in the broadband mode. These two triangles with a similar geometry allow
the direct comparison of structure effects between the legacy VLBI system and the
VGOS system. The VGOS triangle observed only in CONT17 and the S/X triangle observed in
40 sessions in 2018 and 2019. The closure delays with absolute magnitudes larger than
150\,ps in the VGOS plot are very likely due to delay jumps instead of source structure
effects directly, which is discussed in subsection \ref{sec:3.3}. Two solid horizontal lines with an absolute value of 150\,ps are provided as guides.}
\label{fig:3C418}       % Give a unique label
\end{figure}

\paragraph{3C418} Source 3C418 is a representative of the extremely extended
sources in geodetic VLBI and has been observed frequently in the VGOS
sessions. Closure delays for triangle
\texttt{ISHIOKA}--\texttt{KOKEE12M}--\texttt{WETTZ13S} are shown in the bottom
plot of Fig. \ref{fig:3C418}. With replaceable S/X and broadband receivers
at the \texttt{ISHIOKA} station and co-located S/X VLBI stations at the sites
of both \texttt{KOKEE12M} and \texttt{WETTZ13S}, it is possible to have a
similar triangle of stations observing in the S/X mode. Closure delays at X-band 
from the IVS S/X
observations\footnote{https://cddis.nasa.gov/archive/vlbi/ivsdata/vgosdb/} in
2018 and 2019 for triangle \texttt{ISHIOKA}--\texttt{KOKEE}--\texttt{WETTZELL}
were calculated and are shown in the top of the figure. Since the source structure
effects in VGOS delays are due to the structure at the four frequency
bands in the range over 3.0\,GHz to 10.7\,GHz in a complex manner and those in the X-band observations
are due to structure at the frequencies around 8.4\,GHz, the variation patterns in these two plots do not 
necessarily match with each other. However, the scatters of the
closure delays along the variable curves, indicating the random measurement
noise level, are far smaller for VGOS observations than for the S/X
observations. And even for an extended source like 3C418, those scatters for
VGOS observations are at the level of just a few picoseconds. In the bottom
plot, the closure delays with absolute magnitudes larger than 150\,ps are very
likely due to the jumps instead of source structure effects in the delay observables. 
The delay jump issue is discussed further in the next
subsection.

% For one-column wide figures use

\subsection{Delay jumps}
\label{sec:3.3}

In the S/X VLBI mode, multi-band group delay observables have
ambiguities, typically with spacings of 50\,ns (1\,ns\,=\,10$^{-9}$\,s) at X-band
and 100\,ns at S-band, while the VGOS broadband delays have an ambiguity spacing of 31.25\,ns; they can usually be resolved based on a priori information
prior to performing a geodetic VLBI solution. In the broadband VGOS observations reported here, 
jumps in group delays have been found to be at least two orders of
magnitude smaller than the ambiguity spacing of S/X observations, but only
2--3 times the ambiguity spacing of phase delay at X-band. These delay jumps exist in all of the VGOS sessions.

\begin{figure*}[tbhp!]
  \includegraphics[width=0.75\textwidth]{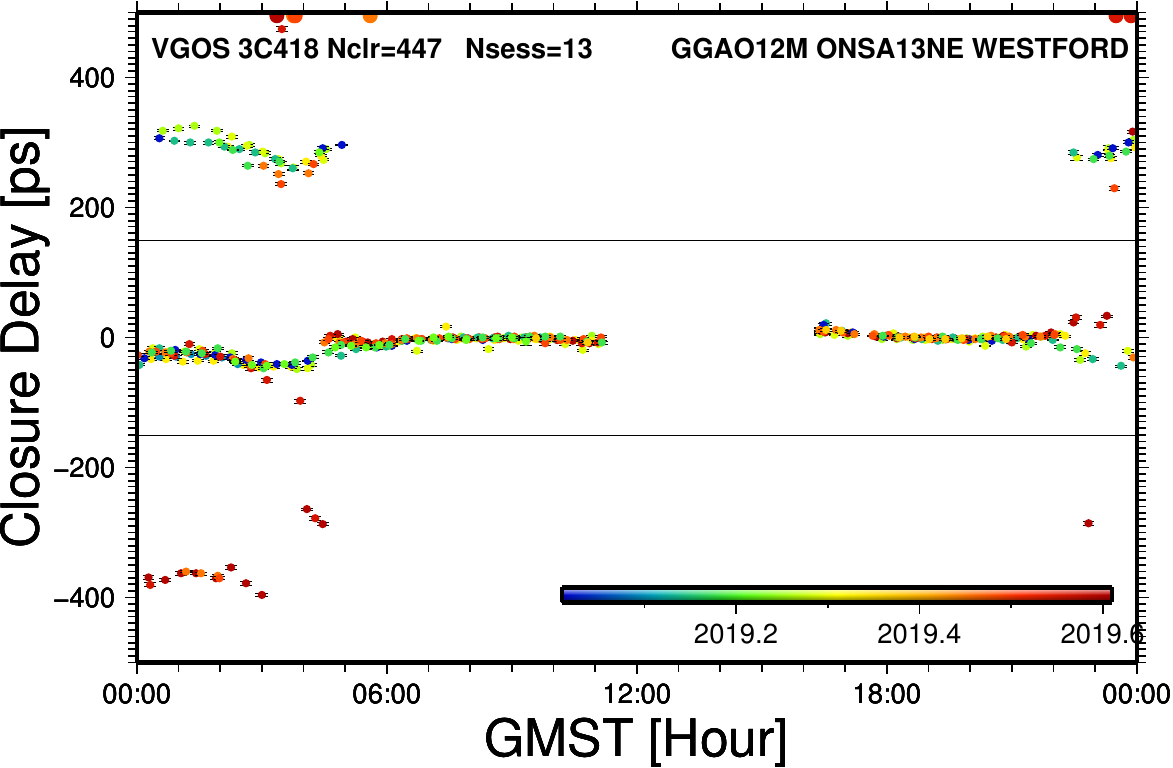}
  \includegraphics[width=0.75\textwidth]{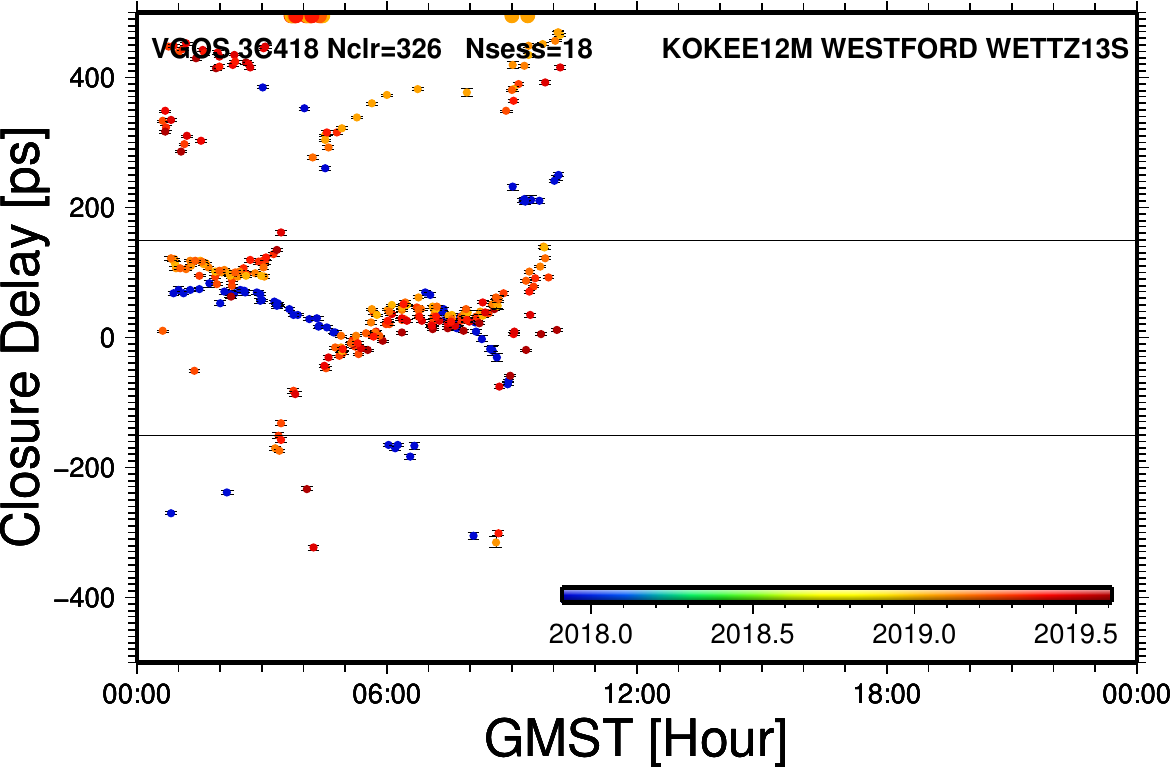}
%  \includegraphics[width=0.75\textwidth]{observations/all_delay_VGOS_3C418_ONSA13NE.pdf}
% figure caption is below the figure
\caption{Closure delays for source 3C418 as a function of GMST for two triangles,
\texttt{GGAO12M}--\texttt{ONSA13NE}--\texttt{WESTFORD} (top) and
\texttt{KOKEE12M}--\texttt{WESTFORD}--\texttt{WETTZ13S} (bottom). For comparison, closure delays of the
second triangle can be seen for sources 0528+483, 0059$+$581 and 0016$+$731 in Figs. \ref{fig:0529+483},
\ref{fig:0059+581} and \ref{fig:0016+731}, respectively.}
\label{fig:3C418_jump}       % Give a unique label
\end{figure*}

Closure delays for 3C418 are shown in Fig. \ref{fig:3C418_jump} for two
triangles, \texttt{GGAO12M}--\texttt{ONSA13NE}--\texttt{WESTFORD} and
\texttt{KOKEE12M}--\texttt{WESTFORD}--\texttt{WETTZ13S}. For the first
triangle, offsets with a magnitude of $\sim$310\,ps occurred
during the time period of GMST 22:00 to 05:00 in 13 VGOS sessions. Even more
complicated delay jumps appear in triangle
\texttt{KOKEE12M}--\texttt{WESTFORD}--\texttt{WETTZ13S}, but no such jumps
show up in the two bottom plots of Figs. \ref{fig:0059+581} and
\ref{fig:0016+731} for 0059$+$581 and 0016$+$731, which cover the same
triangle. These delay jumps are more easily identified in a plot of closure
delays versus closure TEC as shown in Figs. \ref{fig:iono3C418} and
\ref{fig:ionoall}. They also happen frequently for other extended
sources such as 0119$+$115 (CARMS=0.39) and 0229$+$131 (CARMS=0.61). As
demonstrated in Figure 3 of \citet{cappallo2016}, which shows the
two-dimensional fringe amplitudes as a function of $\delta$TEC and group
delay, one would expect big jumps in $\delta$TEC and in group delay if the wrong peak is mistakenly picked up. Since these jumps tend to happen in the case of extended sources and only a few tens of closure delays and closure $\delta$TEC for the CARMS-0.25 sources have jumps, it is likely that the causative factor is source structure. Nevertheless, other reasons are possible as well, for instance, the phasecal problem as found in session VT9007. The sizes of the jumps identified in closure delays seem to be rather stable; however, further studies are necessary to verify if they have a fixed spacing or at what level they can change.

%
%
%% For two-column wide figures use
%\begin{figure*}
%% Use the relevant command to insert your figure file.
%% For example, with the graphicx package use
%  \includegraphics[width=0.75\textwidth]{observations/baseline_err.pdf}
%% figure caption is below the figure
%\caption{Please write your figure caption here}
%\label{fig:2}       % Give a unique label
%\end{figure*}

\section{Ionospheric effects determined by VGOS}

The investigation of $\delta$TEC observables in VGOS is interesting because
(1) unlike the S/X VLBI system, the design of the VGOS system requires that
the dispersion constant in the phase be determined simultaneously with the group delay, and (2)
there is a strong correlation, larger than 0.9, between $\delta$TEC and group
delay estimates based on the current frequency settings, as shown in the
variance-covariance analysis of \citet{cappallo2014, cappallo2016}. Observations on the single baseline \texttt{ISHIOKA}--\texttt{KASHIM34} in
\citet{kondo2016rs} showed that the standard deviation of the differences
between VGOS $\delta$TEC observables and the global TEC model was 0.25\,TECU.
Even though the baseline length of \texttt{KASHIM34}--\texttt{ISHIOKA} (about
50\,km) is too short to make a solid conclusion, the differences are far beyond
the formal errors of VGOS $\delta$TEC observables. The observations of the
single baseline \texttt{GGAO12M}--\texttt{WESTFORD} in \citet{niell2018rs}
showed a consistency between the VGOS $\delta$TEC observables and differenced GNSS TEC
estimates at co-located sites at the level of 1\,TECU. A bias of GPS
relative to VLBI of $-$0.5\,$\pm$\,0.1\,TECU was found in the observations on
this 600\,km baseline. However, neither of these two studies investigated the 
accuracy of the GNSS-based $\delta$TEC used for comparison; consequently, it is 
not clear if these differences come from the VGOS $\delta$TEC estimates or not. The accuracy of, 
and the potential biases in, VGOS $\delta$TEC estimates need to be better understood.

The WRMS $\delta$TEC errors are seen in Table
\ref{tab:iono} to be in the range 0.24\,TECU to 0.49\,TECU for the 20 sessions
excluding VT9007, for which the WRMS error value is 0.73\,TECU. Excluding sessions VT9007 and VT9022, the WRMS $\delta$TEC errors, labelled as ``ALL-19", are 0.31\,TECU to 0.34\,TECU for the two weighting schemes. They are about one order of magnitude larger than the uncertainties of the $\delta$TEC
observables, which implies that there are additional error sources in the
$\delta$TEC observables. The closure analysis of observations of individual sources 
showed that those additional errors in
$\delta$TEC are source-dependent. The WRMS $\delta$TEC error of the
observations for the sources with minimum structure (the CARMS-0.25 group) is
only 0.07\,TECU based on the natural weighting scheme. Source structure
must therefore play a crucial role in the $\delta$TEC measurements.

% For tables use
\begin{table}[tbhp!]
% table caption is above the table
\caption{WRMS $\delta$TEC errors determined by closure analysis (in units of TECU)}
\label{tab:iono}       % Give a unique label
% For LaTeX tables use
\begin{tabular}{lrr}
\hline\noalign{\smallskip}
Session/Group & Uniform Weighting & Natural Weighting\\
(1) & (2) & (3) \\
\noalign{\smallskip}\hline\noalign{\smallskip}
B17337& 0.34& 0.32\\
B17338& 0.45& 0.34\\
B17339& 0.32& 0.30\\
B17340& 0.37& 0.37\\
B17341& 0.35& 0.34\\
VT9007& 0.56& 0.73\\
VT9022& 0.33& 0.29\\
VT9035& 0.28& 0.24\\
VT9050& 0.28& 0.26\\
VT9063& 0.32& 0.30\\
VT9077& 0.33& 0.28\\
VT9091& 0.33& 0.32\\
VT9105& 0.32& 0.30\\
VT9119& 0.32& 0.38\\
VT9133& 0.33& 0.27\\
VT9148& 0.33& 0.37\\
VT9162& 0.36& 0.28\\
VT9175& 0.34& 0.24\\
VT9189& 0.39& 0.49\\
VT9203& 0.34& 0.37\\
VT9217& 0.40& 0.32\\
ALL&    0.36& 0.35\\
\textbf{ALL-19}& \textbf{0.34}&  \textbf{0.31}\\
\textbf{CARMS-0.25}& \textbf{0.15} & \textbf{0.07}\\
\noalign{\smallskip}\hline
\end{tabular}{}
\end{table}

\section{Correlation between $\delta$TEC and group delay observables from VGOS}

\begin{figure*}[tbhp!]
  \includegraphics[width=0.75\textwidth]{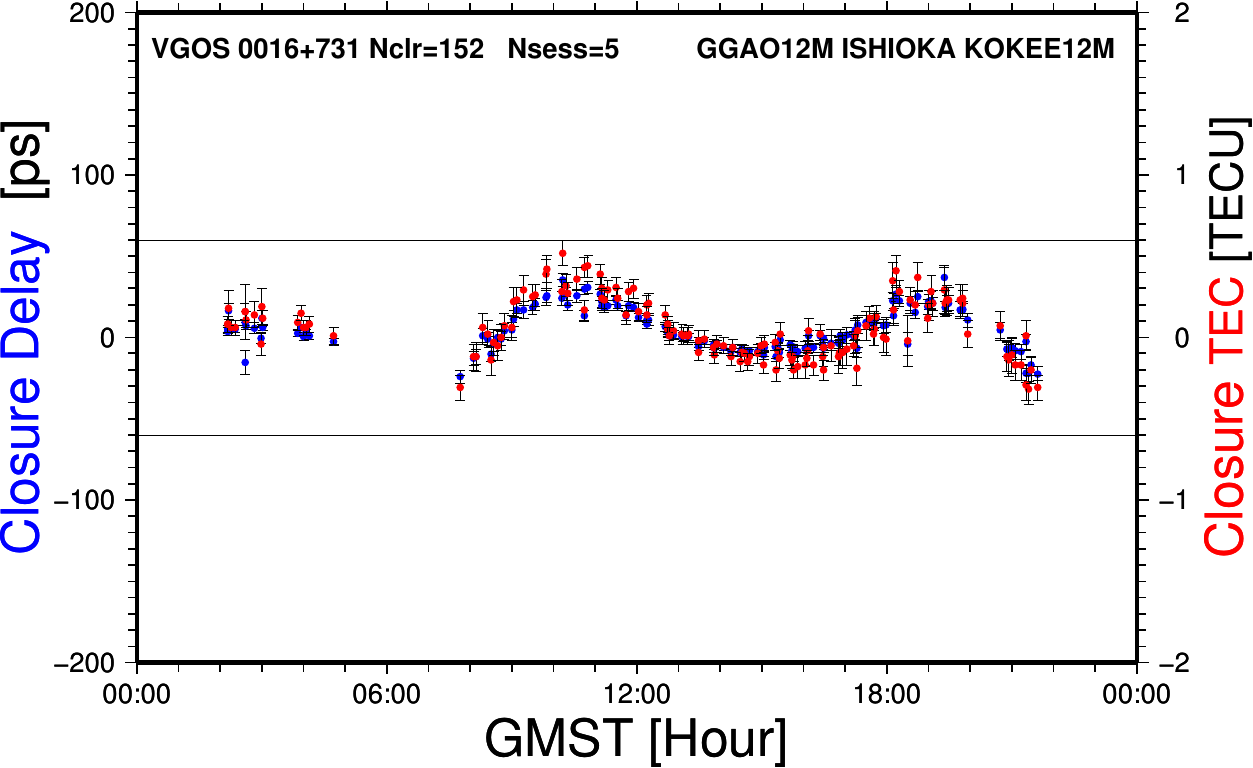}
  \includegraphics[width=0.69\textwidth]{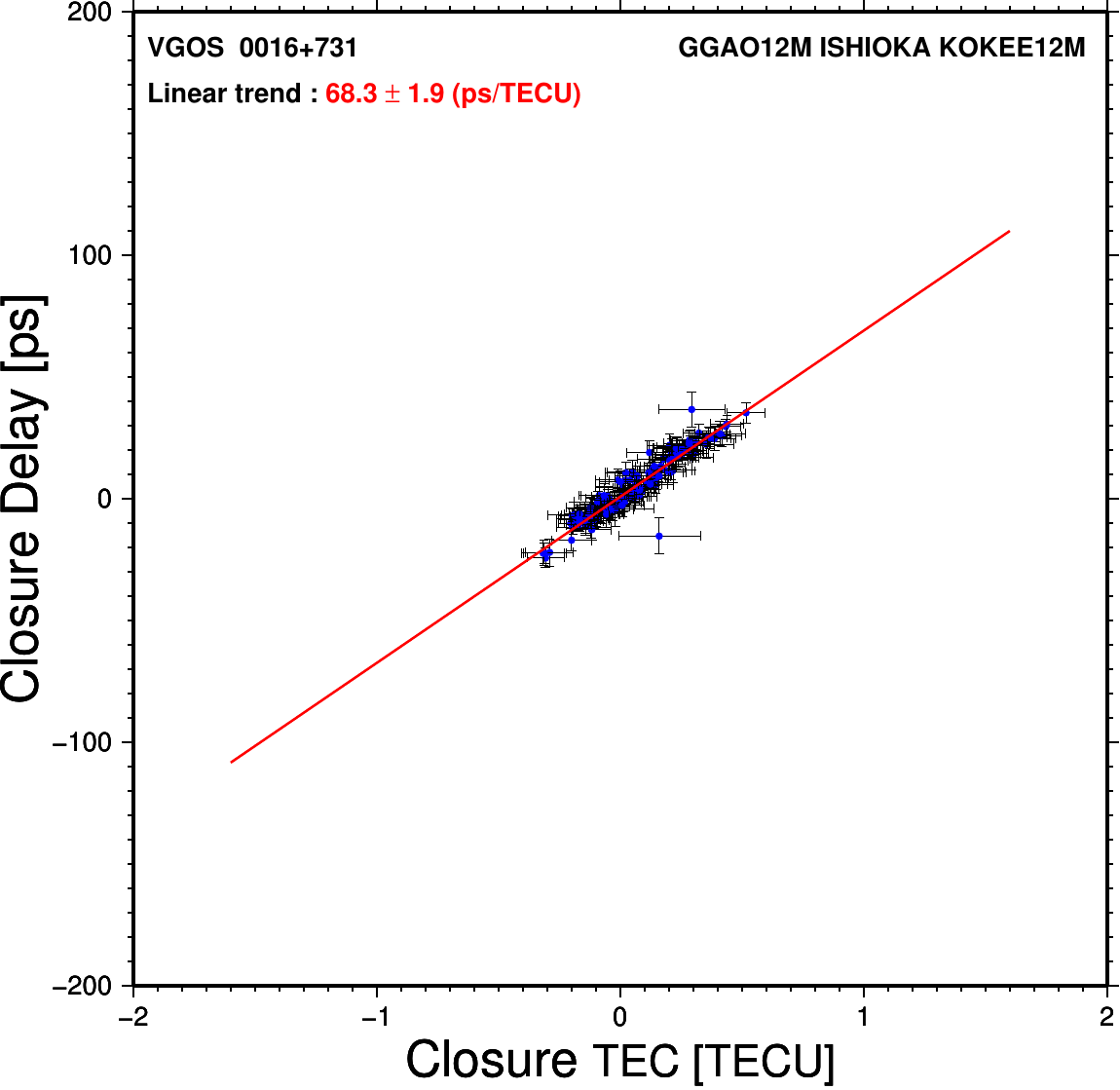}
% figure caption is below the figure
\caption{Demonstration of the strong correlation between $\delta$TEC and group delay observables from VGOS.
Closure delays (blue dots) and closure TEC (red dots) for source 0016$+$731 for triangle
\texttt{GGAO12M}--\texttt{ISHIOKA}--\texttt{KOKEE12M} as a function of GMST are shown in the
top plot, whereas these closure delays versus closure TEC are in the bottom plot.
The changing pattern in closure TEC is the same as that of closure delays. There is a strong 
correlation between them, and the linear trend is 68.3\,$\pm$\,1.9\,ps/TECU.}
\label{fig:TEC0016+731}       % Give a unique label
\end{figure*}

\begin{figure*}[tbhp!]
  \includegraphics[width=0.75\textwidth]{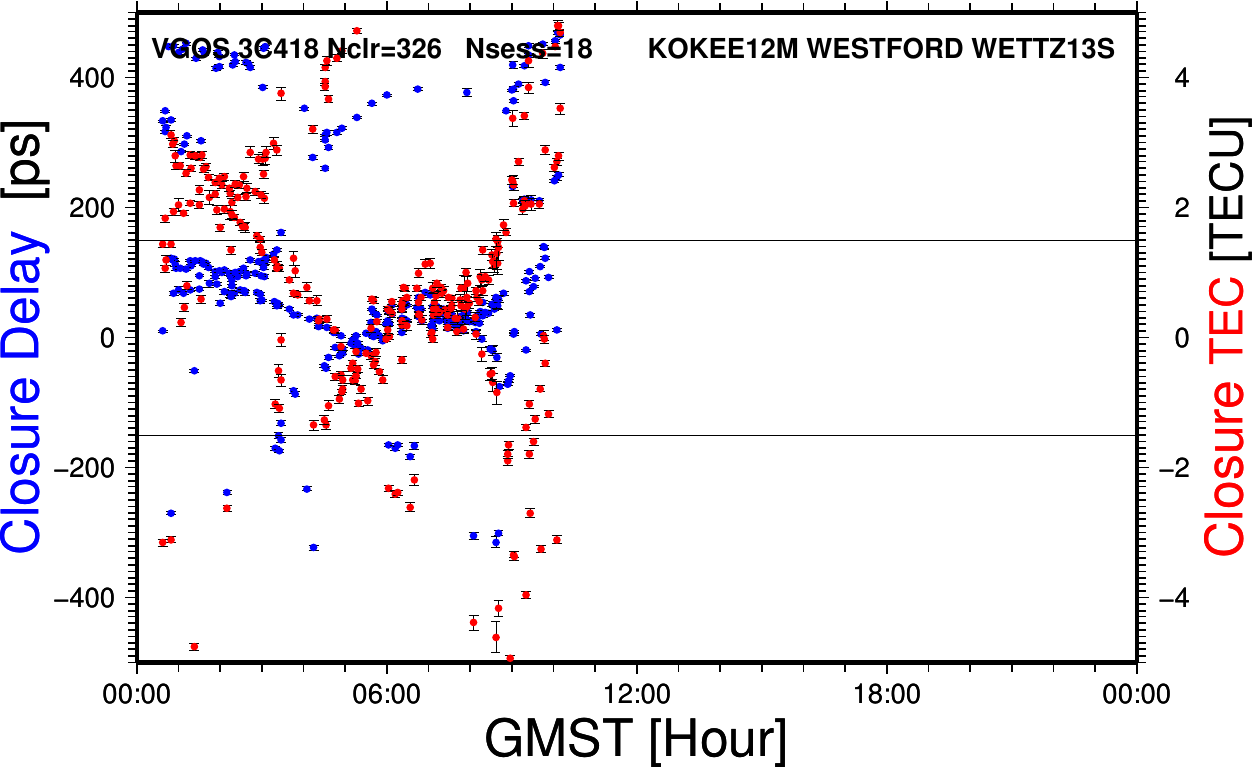}
  \includegraphics[width=0.69\textwidth]{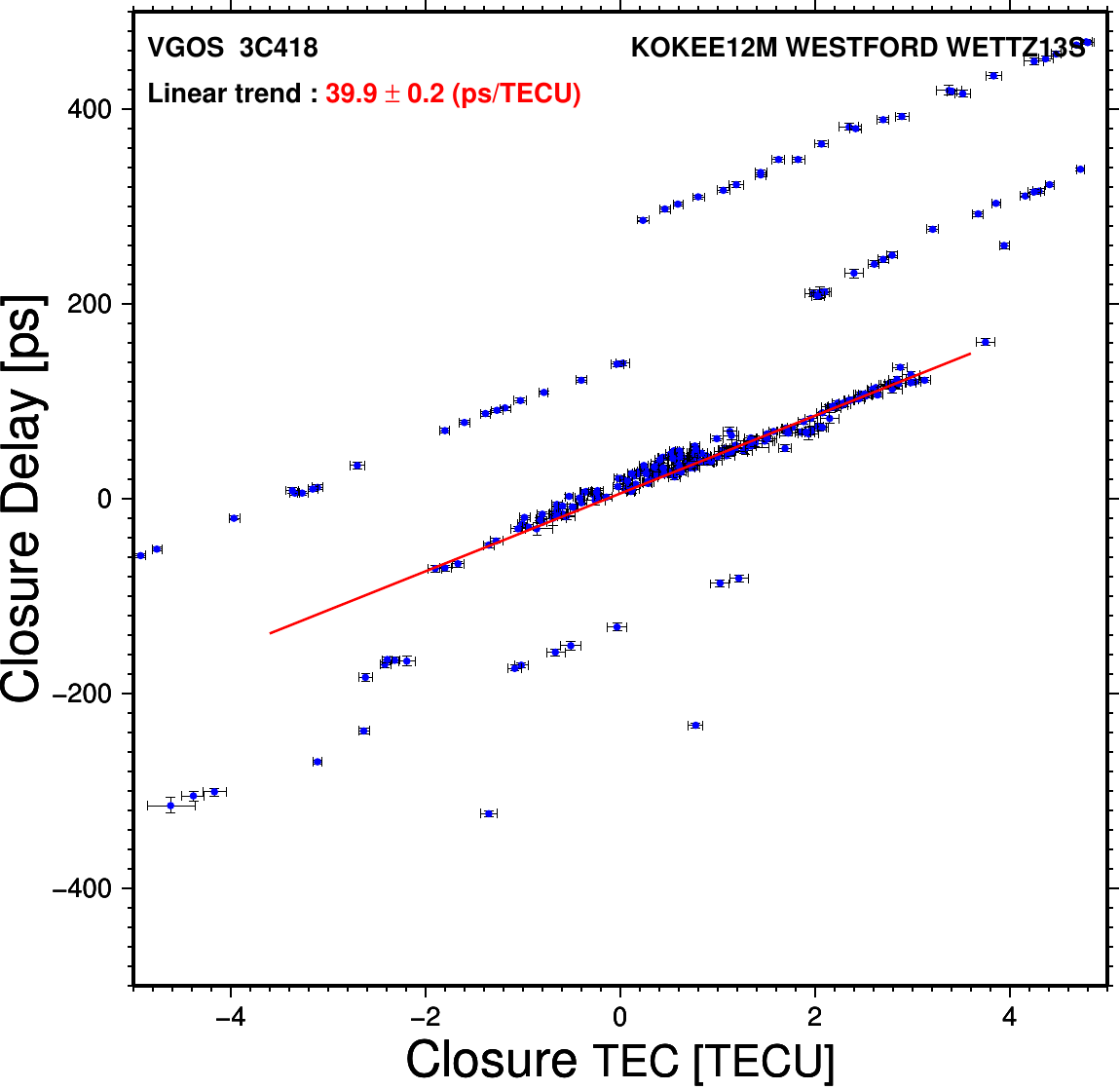}
% figure caption is below the figure
\caption{Demonstration of the strong correlation between $\delta$TEC and group delay observables
from VGOS and the jumps in them.
Closure delays (blue dots) and closure TEC (red dots) for source 3C418 for triangle
\texttt{GGAO12M}--\texttt{KOKEE12M}--\texttt{WETTZ13S} as a function of GMST are shown in the
top plot, whereas these closure delays versus closure TEC are in the bottom plot.
The linear trend between them is 39.9\,$\pm$\,0.2\,ps/TECU. The jumps,
which can be two times a certain interval away from the mainstream of points passing the zero closure delay and zero closure TEC}, are clearly visible in the bottom plot.
\label{fig:iono3C418}       % Give a unique label
\end{figure*}

A covariance analysis using the VGOS frequency setup predicts 
a strong correlation between the group delay and $\delta$TEC
estimates \citep[see][]{cappallo2015}. It can be more straightforward to understand that correlation and
its influence on VGOS observations by analyzing the actual data.
Figures \ref{fig:TEC0016+731} and \ref{fig:iono3C418} demonstrate the
correlation by showing closure delays and closure TECs for the sources 0016$+$731
and 3C418 using two plots each. The trends, obtained from least-square fitting (LSQ), are 68.3\,$\pm$\,1.9\,ps/TECU and 39.9\,$\pm$\,0.2\,ps/TECU for the two sources, 
respectively. 

In the bottom plot of Fig. \ref{fig:iono3C418}, the points deviating
significantly from the red line form basically four straight lines that are
parallel to the red line with offsets of 133\,ps in delay or 3.3\,TECU
in $\delta$TEC from each other. It confirms the jumps in either or both the group delay
and $\delta$TEC observables. 

\begin{figure*}[tbhp!]
\includegraphics[width=0.7\textwidth]{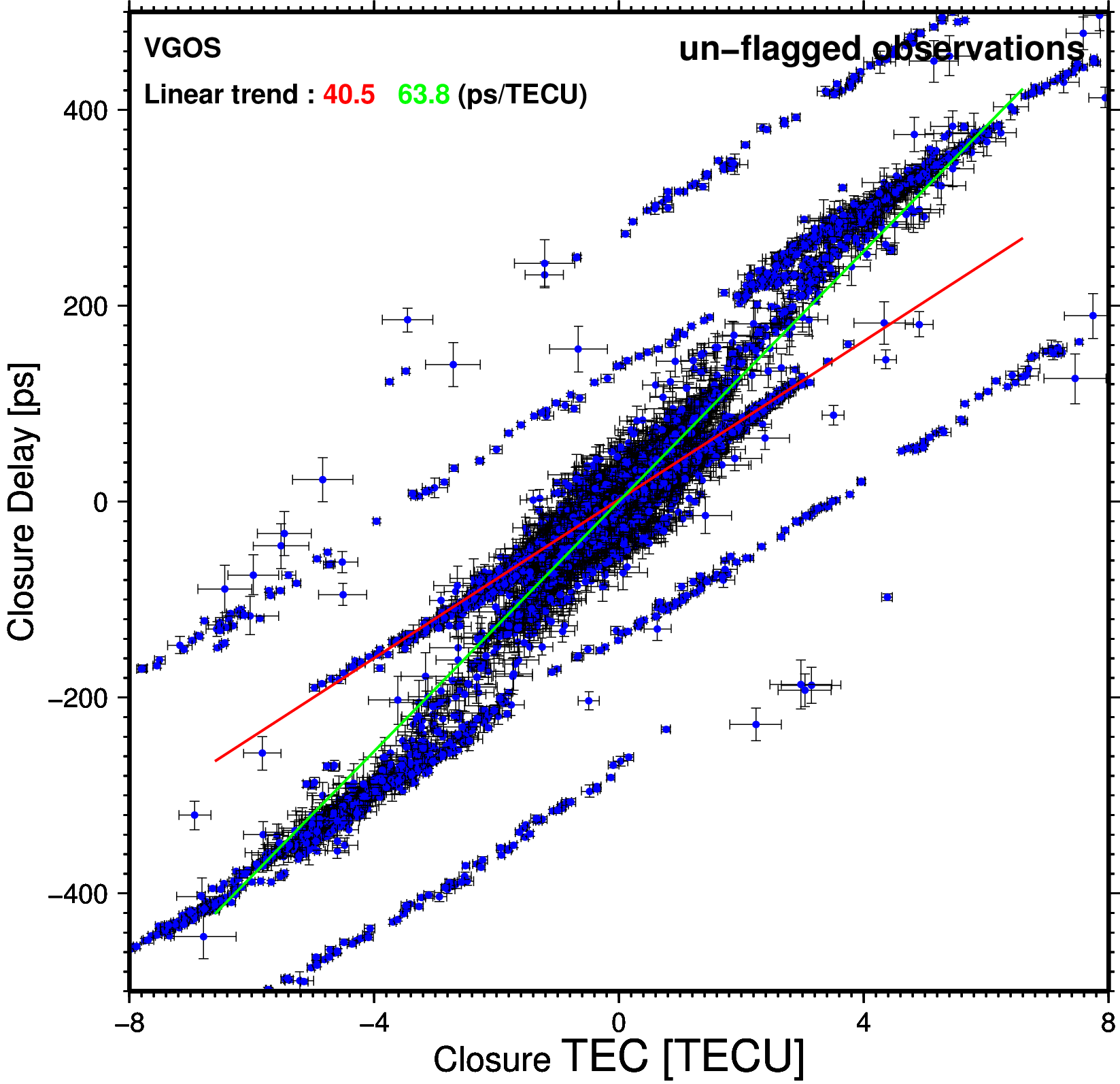}
\includegraphics[width=0.7\textwidth]{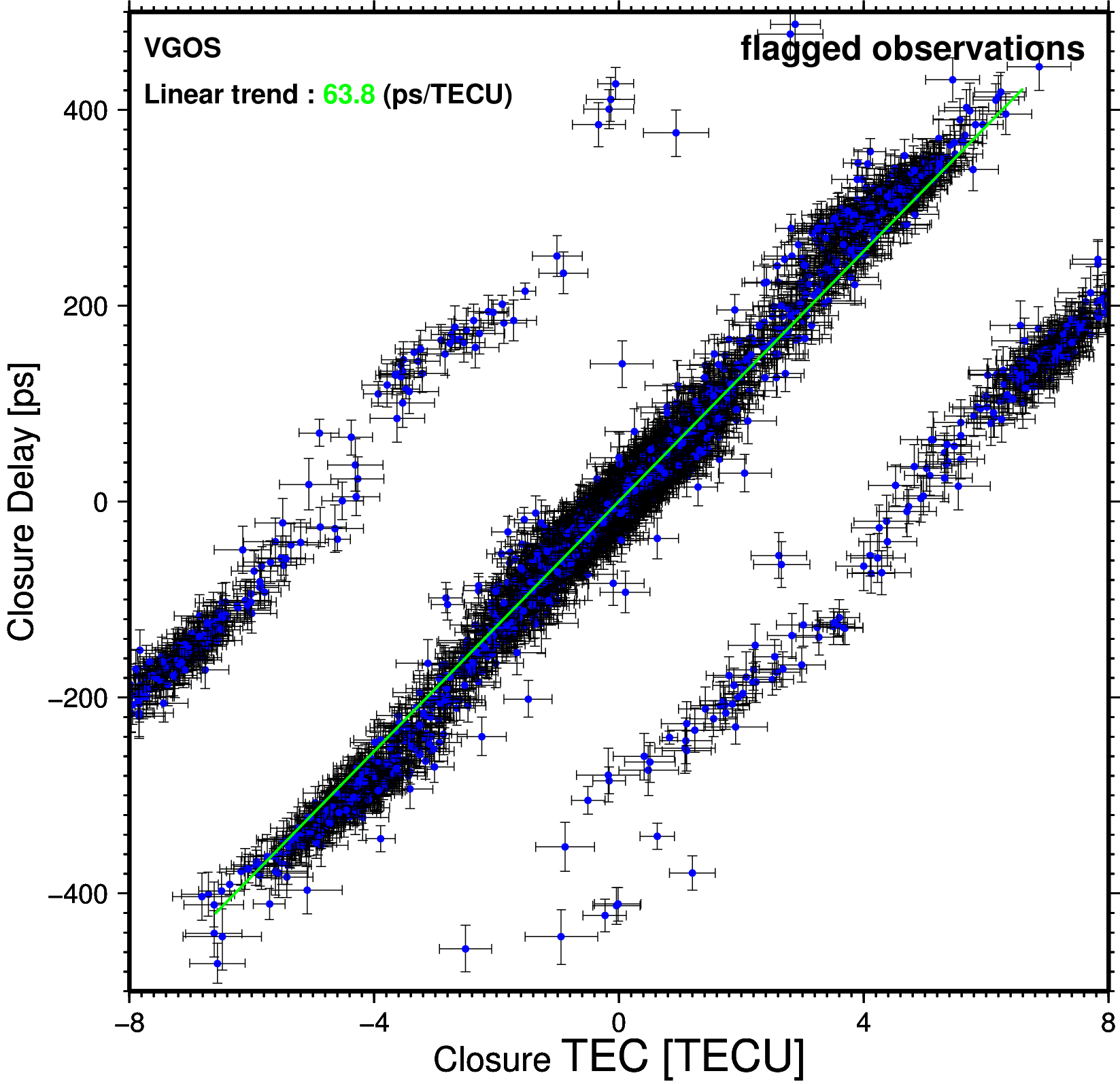}
% figure caption is below the figure
\caption{Closure delays versus closure TEC with un-flagged observations (top) and with at least one of the three observations in a triangle flagged out due to the three cases listed in section \ref{section_data} (bottom).  All sources and all triangles available in the 21 sessions are included. There are two main linear trends between them. 
The slope of the trend indicated by the red line was determined to be 40.5\,$\pm$\,0.1\,ps/TECU. That of the trend indicated by the green line with a slope of 63.8\,ps/TECU was found to be in the range of 59.9\,ps/TECU to 68.9\,ps/TECU depending on the flagging and weighting scheme, which indicates that it is variable from source to source and from triangle to triangle. In the bottom plot, the vast majority of the closures are from observations of station \texttt{RAEGYEB} in the last four sessions in CONT17; while the closures of the observations flagged due to the other two reasons are nearly all beyond the limits of the plotting axes. The observations of station \texttt{RAEGYEB} have a median SNR of 17-23 in these four sessions, while the rest observations in CONT17 have a median SNR of 92--115. On average, the closures in the bottom plot are from observations with SNRs smaller by a factor of five than those in the upper plot. Another difference in the observations between the two plots is the significant decrease in the channel visibility amplitudes of station \texttt{RAEGYEB} with increasing frequency due to the antenna pointing issue since the second day during the CONT17.}
\label{fig:ionoall}       % Give a unique label
\end{figure*}

Figure \ref{fig:ionoall} shows the closure delays as a function of the closure TEC for all sources and all triangles in the 21 sessions. The closure quantities in the upper plot are from un-flagged observations, whereas those in the bottom plot have at least one of the three observations in a triangle flagged due to the three cases listed in section \ref{section_data}. Two main linear trends between closure delay and $\delta$TEC were identified. In the upper plot the data points grouped in the lines parallel to the red line were used jointly to determine the slope with a result of 40.5\,$\pm$\,0.1\,ps/TECU. This estimated value is different from that derived from the closure quantities shown in Fig.~\ref{fig:iono3C418} for source 3C418 by three times their derived uncertainties, suggesting that the uncertainties from LSQ were too optimistic. Based on the remaining data, another linear trend as indicated by the green line with a slope of 63.8\,ps/TECU was determined and found to be in the range of 59.5\,ps/TECU to 68.9\,ps/TECU depending on the flagging and weighting schemes. The results of these two trends were iteratively determined by excluding the data points larger than five times the WRMS residual. These two linear trends seem to have different origins: (1) the trend in the range 59.9\,ps/TECU to 68.9\,ps/TECU agrees with the value of $\sim$62\,ps/TECU from \citet{cappallo2016} and is due to the random measurement noise in the channel phases across the four bands; (2) the trend of $\sim$40\,ps/TECU results from the systematic variations in the channel phases due to source structure.

\begin{figure*}[tbhp!]
\includegraphics[width=0.75\textwidth]{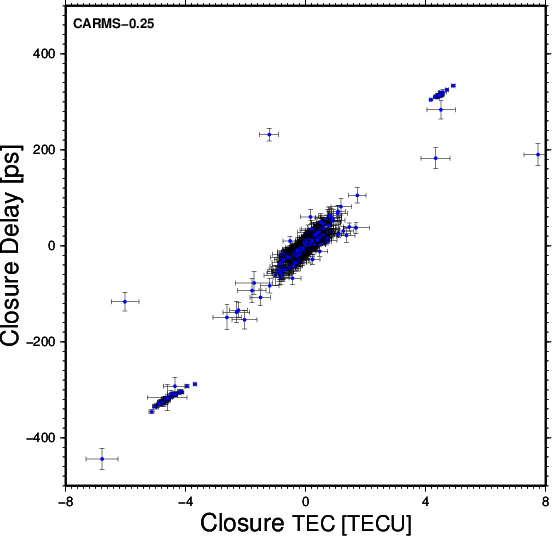}
% figure caption is below the figure
\caption{Equivalent plot to Fig. \ref{fig:ionoall} for the closure quantities of the CARMS-0.25 observations only. Of 20,337 pairs of closure quantities in the plot, there are 17 and 47 pairs in the upper-right and the bottom-left corners, respectively. Most of them involve the observations of sources 0133$+$476 and 0716$+$714. The median value of the absolute closure delays in the plot is 2.38\,ps, and that of the absolute closure TEC is 0.059\,TECU.}
\label{fig:carms025}       % Give a unique label
\end{figure*}

Figure \ref{fig:carms025} is an equivalent plot for CARMS-0.25 sources. Other than the small isolated groups of closures in the upper right and lower left, which are associated primarily with only two of the 28
sources in this category, there are no jumps comparable to those seen in Fig. \ref{fig:ionoall}. Were the points
for the CARMS-0.25 sources removed, the jumps would still be prevalent. Since the closures
shown in Fig. \ref{fig:ionoall} are for all sources, removing the points for the CARMS-0.25 sources would leave
the closures for the sources with CARMS greater than 0.25; these are the sources with nominally the
more extended source structure. Therefore, our findings indicate that the predominant causative factor of the jumps in delay and $\delta$TEC is source structure, which can cause large frequency-dependent phase variations across the four bands. This has been demonstrated by our recent imaging results based on closure phases and closure amplitudes \citep[figures 11 and 12 in][]{10.1002/essoar.10504599.1}.

%
%%
%\begin{figure*}
%  \includegraphics[width=0.75\textwidth]{observations/all_di_VGOS_0016+731_fig9.pdf}
%  \includegraphics[width=0.75\textwidth]{observations/all_id_VGOS_0016+731_fig9.pdf}
%% figure caption is below the figure
%\caption{Please write your figure caption here}
%\label{fig:2}       % Give a unique label
%\end{figure*}
%%%
%\begin{figure*}
%  \includegraphics[width=0.75\textwidth]{observations/all_di_VGOS_3C371_fig28.pdf}
%  \includegraphics[width=0.75\textwidth]{observations/all_id_VGOS_3C371_fig28.pdf}
%%% figure caption is below the figure
%\caption{Please write your figure caption here}
%\label{fig:2}       % Give a unique label
%\end{figure*}
%%%
%
%\begin{figure*}
%  \includegraphics[width=0.75\textwidth]{observations/all_di_VGOS_3C371_fig39.pdf}
%  \includegraphics[width=0.75\textwidth]{observations/all_id_VGOS_3C371_fig39.pdf}
% figure caption is below the figure
%\caption{Please write your figure caption here}
%\label{fig:2}       % Give a unique label
%\end{figure*}
%
%%%%
%\begin{figure*}
%  \includegraphics[width=0.75\textwidth]{observations/all_di_VGOS_3C418_fig9.pdf}
%  \includegraphics[width=0.75\textwidth]{observations/all_id_VGOS_3C418_fig9.pdf}
% figure caption is below the figure
%\caption{Please write your figure caption here}
%\label{fig:2}       % Give a unique label
%\end{figure*}

%%
\section{Conclusions and discussions}

We processed the 21 VGOS sessions that have been publicly released and made 
quality assessments for two kinds of VGOS observables, group delay and
$\delta$TEC, that are determined simultaneously in the process of broadband
bandwidth synthesis. The measurement noise level and the contributions of
systematic error sources in these two types of observables were determined by
running closure analysis for the whole data set and for the selected sources
with minimum structure based on our previous work. By performing closure
analysis, two important features in group delay and $\delta$TEC observables
have been revealed, which are the strong correlation between 
them and the jumps in both observables.

The random measurement noise level of VGOS group delays was found to be below
2\,ps based on the observations from all the VGOS radio sources that have
CARMS values smaller than 0.25. The estimated random measurement noise level
agrees well with the delay formal errors, as listed in Table \ref{tab:1}.
However, the contributions from other systematic error sources, mainly source
structure related, are at the level of 20\,ps, as indicated by the WRMS delay errors for observations of all sources. 
Due to the significant reduction in measurement
noise over the S/X systems, source structure effects with magnitudes of 10\,ps are clearly visible.
In general, source structure evolves at time scales of a few weeks, which causes the closure delays to change at magnitudes of a few tens of picoseconds. It thus will be a big challenge to correct source structure effects in VGOS in order to fulfill its goals. Evidence for another critical error source in the VGOS system
is the presence of discrete jumps in the closure delays and closure TECs, for instance with a delay offset of about 
310\,ps or integer multiples of that. The likely cause is found to be source-structure-induced phase changes across the four bands \citep{10.1002/essoar.10504599.1}.

Closure delays on individual triangles were shown for four sources, 0529$+$483 (CARMS=0.21), 0059$+$581 (CARMS=0.27), 0016+731 (CARMS=0.31), and 3C418 (CARMS=0.61) in figures \ref{fig:0529+483}, \ref{fig:0059+581}, \ref{fig:0016+731}, and \ref{fig:3C418_jump} to demonstrate the source structure effects in VGOS delay observables. By showing the closure delays from the same triangle \texttt{KOKEE12M}--\texttt{WESTFORD}--\texttt{WETTZ13S} 
in these four figures, the differences in the magnitudes of these effects can be compared among radio sources with structure at various scales as indicated by their CARMS values. The magnitudes of structure effects on the triangle were less than 10\,ps for source 0529+483, about\,50 ps for source 0059+581, and about 100\,ps for source 0016+731; they were larger and more complicated for source 3C418. Delay jumps occurred for the observations of only 3C418 among these four sources.

The random measurement noise level of $\delta$TEC observables was determined to be
below about 0.07\,TECU, which is comparable to the formal errors. The
systematic effects are five times larger than that. A strong correlation
between group delay and $\delta$TEC observables is clearly demonstrated, with
two main linear trends. For observations with large structure effects, there is 
a dominant slope of $\sim$40\,ps/TECU. The slope of the second trend is 
in the range 60\,ps/TECU to 70\,ps/TECU. 
Due to this strong correlation and the simultaneous determination of them, 
group delay and $\delta$TEC observables 
need to be studied together and further. The $\delta$TEC
estimates from other sources, such as GPS or global TEC models with a sufficient
accuracy, might improve the determination of the source structure effects in $\delta$TEC
observables; based on the stable linear coefficients between delay and
$\delta$TEC, the source structure effects in group delay observables might be determined
without requiring any model of source structure itself. For example, external $\delta$TEC estimates can be used to detect the systematic effects in VGOS $\delta$TEC estimates, which may be able to predict those effects in delay observables by the linear trends, as discussed in this work.

Delay jumps in the VGOS system need to be understood further. Closure delays have
been demonstrated to be useful, and the correlation between group delay and
$\delta$TEC observables can also be of great help for the delay jump detection.
However, the delay spacing of these jumps will have to be studied in
detail. The exact origins of the two dominant linear trends between broadband delays and $\delta$TEC, the causes of such jumps, and the method to fix them
are our near-future work.

\begin{acknowledgements}
%If you'd like to thank anyone, place your comments here
%and remove the percent signs.
We would like to thank Sergei Bolotin, Arthur Niell, and Brian Corey for their efforts to review the manuscript and for their helpful comments of high quality which improved its comprehensibility. The results reported in this paper were produced using the data owned by the International VLBI Service (IVS) and its international self-funded member organizations. We are grateful to the IVS VGOS stations at GGAO (MIT Haystack Observatory and NASA GSFC, USA), Ishioka (Geospatial Information Authority of Japan), Kokee Park (U.S. Naval Observatory and NASA GSFC, USA), Onsala (Onsala Space Observatory, Chalmers University of Technology, Sweden), Westford (MIT Haystack Observatory), Wettzell (Bundesamt f\"{u}r Kartographie und Geod\"{a}sie and Technische Universit\"{a}t M\"{u}nchen, Germany), and Yebes (Instituto Geogr\'{a}fico Nacional, Spain), to the staff at the MPIfR/BKG correlator center, the VLBA correlator at Socorro, and the MIT Haystack Observatory correlator for performing the correlations and the fringe fitting of the data, and to the IVS Data Centers at BKG (Leipzig, Germany), Observatoire de Paris (France), and NASA CDDIS (Greenbelt, MD, USA) for the central data holds.\\
This research has made use of the Generic Mapping Tools package\footnote{https://www.soest.hawaii.edu/gmt/}, the pgplot library\footnote{https://www.astro.caltech.edu/$\sim$tjp/pgplot/}, and the SAO/NASA Astrophysics Data System\footnote{https://ui.adsabs.harvard.edu/}.\\ MHX was supported by the Academy of
Finland project No. 315721 and by the National Natural Science Foundation of China (No. 11973023 and 11873077).
JMA, RH, SL, and HS were supported by the German Research
Foundation grants HE5937/2-2 and SCHU1103/7-2.
\end{acknowledgements}

\section*{Author contributions}
MHX and JMA designed the research; MHX performed the research, analyzed the data, and wrote the paper; JMA, RH, SL, HS, and GW contributed to the interpretation of the results and provided suggestions in writing and revising the paper. 

\section*{Data Availability}
All the VGOS data in the vgosDB data format are available at the CDDIS server: https://cddis.nasa.gov/archive/vlbi/ivsdata/vgosdb/. The closure
delays, closure TECs, closure phases and amplitudes from these VGOS sessions
are available upon the request to the corresponding author.

% Authors must disclose all relationships or interests that
% could have direct or potential influence or impart bias on
% the work:
%
% \section*{Conflict of interest}
%
% The authors declare that they have no conflict of interest.
% BibTeX users please use one of
\bibliographystyle{spbasic}      % basic style, author-year citations
\bibliography{export-bibtex}   % name your BibTeX data base

% Non-BibTeX users please use
%\begin{thebibliography}{}
%%
%\bibitem[Anderson \& Xu(2018a)]{and18} Anderson, J. M. \& Xu, M. H., 2018a,
%    JGR, https://doi.org/10.1029/2018JB015550
%\bibitem[Fey et al.(2015)]{fey15} Fey, A., Gordon, D. ,Jacobs C. S., et al.,
%    2015, AJ, 150, 58
%\bibitem[Niell et al.(2006)]{nie06} Niell, A. E., Whitney A. R., Petrachenko,
%    B., et al., 2006, in 2005 IVS Annual Report NASA/TP-2006-214136,
%    https://ivscc.gsfc.nasa.gov/publications/ar2005/spcl-vlbi2010.pdf
%\bibitem[Niell et al.(2018)]{nie18} Niell, A. E., Barrett J., Burns A., et
%    al., 2018, Radio science, 53, https://doi.org/10.1029/2018RS006617
%
%\bibitem[Nothnagel et al.(2017)]{not17} Nothnagel, A., Artz, T., Behrend, D.,
%    et al., 2017, JG, 91, 711
%\bibitem[Xu et al.(2016)]{xu16} Xu, M. H., Heinkelmann R., Anderson, J. M., et
%    al., 2016, AJ, 152, 151
%
%\bibitem[Xu et al.(2017)]{xu17} Xu, M. H., Heinkelmann R., Anderson, J. M., et
%    al., 2017, JG, 91, 767
%
%
%% etc
%\end{thebibliography}

\end{document}